\pdfoutput=1

\documentclass[11pt,twoside,a4paper,cmspaper,final,collab]{cms-tdr}
\def\svnVersion{b242837-D}\def\svnDate{2019/12/16}\def\cmsCernNoTag{CERN-EP-2018-320}\def\cmsCernDate{\today}\def\cmsMessage{"Published in the Journal of High Energy Physics as \href{http://dx.doi.org/10.1007/JHEP12(2019)059}{\doi{10.1007/JHEP12(2019)059}.}"}

\begin{document}\cmsNoteHeader{SMP-17-001}
\newcommand{\AMCATNLO}{\textsc{MadGraph5\_aMC@NLO}}

\hyphenation{had-ron-i-za-tion}
\hyphenation{cal-or-i-me-ter}
\hyphenation{de-vices}
\RCS$HeadURL$
\RCS$Id$

\providecommand{\PYTHIASIX} {\PYTHIA{}6\xspace}
\providecommand{\PYTHIAEIGHT} {\PYTHIA{}8\xspace}

\cmsNoteHeader{SMP-17-001}
\title{Measurement of the differential Drell--Yan cross section in proton-proton collisions at $\sqrt{s} = 13\TeV$}

\date{\today}

\abstract{
   Measurements of the differential cross section for the Drell--Yan process, based on proton-proton collision data at a centre-of-mass energy of 13\TeV, collected by the CMS experiment, are presented.
   The data correspond to an integrated luminosity of 2.8 (2.3)\fbinv in the dimuon (dielectron) channel.
   The total and fiducial cross section measurements are presented as a function of dilepton invariant mass
   in the range 15 to 3000\GeV, and compared with the perturbative predictions of the standard model.
   The measured differential cross sections are in good agreement with the theoretical calculations.
}

\hypersetup{%
pdfauthor={CMS Collaboration},%
pdftitle={Measurement of the differential Drell-Yan cross section
          in proton-proton collisions at sqrt(s) = 13 TeV},%
pdfsubject={CMS},%
pdfkeywords={CMS, physics, Electroweak, Standard Model, Drell-Yan}}

\maketitle
\newpage
\section{Introduction}

Drell--Yan (DY) production of same-flavour, oppositely charged lepton pairs in proton-proton collisions occurs via the $s$-channel exchange of $\gamma^{*}/$\cPZ~bosons.
Current theoretical predictions for the cross section are accurate up to next-to-leading order (NLO) in the electroweak (EW) coupling and up to next-to-next-to-leading
order (NNLO) in perturbative quantum chromodynamics (QCD)~\cite{QCDNNLO, DYNNLO, DYNNLO1, DY-Theory}. Hence, a precision measurement of the differential cross section for the DY process at the LHC provides an important test of the perturbative framework of the standard model (SM). In a complementary way, the experimental results can also be used to constrain the parton distribution functions (PDFs).
 DY production of dileptons is also a major source of background for rare SM processes as well as searches for physics beyond the SM. Hence, it is important to measure the DY production rate accurately up to the largest accessible energy.

The single- and double-differential cross sections, $\rd\sigma / \rd{m}$ and $\rd^2\sigma / \rd{m} \rd\abs{y}$,
where $m$ is the dilepton invariant mass and $\abs{y}$ is the absolute value of the dilepton rapidity,
were previously measured by the ATLAS~\cite{ATLASDY1, ATLASDY2, ATLASDY3}
and CMS~\cite{CMSDY1, CMSDY2, CMSDY3} Collaborations in proton-proton ({\Pp}{\Pp}) collisions for the centre-of-mass energy $\sqrt{s} = 7$ and 8\TeV at the LHC.
Using early data collected in 2015 at $\sqrt{s} = 13\TeV$ and corresponding to an integrated luminosity (${\cal L}$) of 81\pbinv, the ATLAS Collaboration measured the \cPZ~production cross section times branching ratio of $\cPZ\to \ell^+\ell^-$ near the resonance region~\cite{Aad:2016naf}.
This paper presents the first measurement of the DY spectrum $\rd\sigma / \rd{m}$ over a wider invariant mass range using 2015 {\Pp}{\Pp}~collision data collected by the CMS Collaboration at $\sqrt{s} = 13\TeV$ at the LHC, corresponding to $\mathcal{L} = 2.8$ (2.3)\fbinv in the dimuon (dielectron) channel. The range of $x$, the momentum fraction carried by an interacting parton, covered by this measurement is $10^{-4} < x < 1.0$.

The cross section as a function of invariant mass for a specific bin $i$ is determined by the following relation:
\begin{equation}
\sigma_i = \frac{N_i}{A_i \, \varepsilon_i \, \rho_i \, {\mathcal{L}}},
\label{eqn:eq1}
\end{equation}
where $N_i$ denotes the signal yield in a given bin $i$ after subtracting the background. It is obtained using an unfolding technique to correct for the detector resolution effects and final-state photon radiation (FSR). The acceptance $A_i$ and the experimental efficiency $\varepsilon_i$ are obtained from the Monte Carlo (MC) simulation. A scale factor $\rho_i$ accounts for any difference in the efficiency between data and simulation.

\section{The CMS detector}
\label{sec:det}

The central feature of the CMS apparatus is a superconducting solenoid of 6\unit{m} internal diameter,
providing a magnetic field of 3.8\unit{T}. Within the solenoid volume are a silicon pixel
and strip tracker, a fine-grained and hermetic crystal electromagnetic calorimeter (ECAL), and a sampling
hadron calorimeter, each composed of a barrel and two endcap sections.
Extensive forward calorimetry with lead and quartz-fibre Cherenkov detectors complement the coverage provided by the barrel and endcap detectors. Muons are measured in gas-ionisation detectors using three technologies: drift tubes, cathode strip chambers, and resistive-plate chambers  embedded in the steel flux-return yoke outside the solenoid.

The ECAL consists of about 76k lead tungstate crystals, which provide a coverage in pseudorapidity $\abs{ \eta }< 1.479 $ in the barrel region and $1.566 <\abs{ \eta } < 2.5$ in two endcap regions (EE). A preshower detector consisting of two planes of silicon sensors interleaved with a  total of $3 X_0$ of lead is located in front of EE. The momentum of an electron candidate is estimated by combining the measurements in the ECAL and tracker. The resolution of transverse momentum (\pt) for electron candidate with \pt about 45\GeV, ranges from 1.7\% for nonshowering electrons in the barrel region to 4.5\% for showering electrons in the endcaps~\cite{CMS:EGM-13-001}.

Muons are measured in the range $\abs{\eta}< 2.4$.
Matching the track of a muon candidate to that measured in the silicon tracker results in a
\pt resolution  of 1.3--2.0\% in the
barrel and better than 6\% in the endcaps for muons with $20 <\pt < 100\GeV$.
The  resolution is better than 10\% for muons with \pt up to about 1\TeV~\cite{CMS-PAPER-MUO-16-001}.

The first level of the CMS trigger system~\cite{bib:CMSTrigger}, composed of custom hardware processors, uses information
from the calorimeters and the muon detector to select the most interesting events at a level of 1 in $10^4$
  within 4\mus. The high-level trigger, consisting of processor farms, uses the complete information from
the detector to reconstruct the event and discriminate further to reduce the selection rate to less than
about 1\unit{kHz}, before data storage.
A more general description of the CMS detector with a definition of the coordinate system
used and the relevant kinematic variables, can be found in Ref.~\cite{Chatrchyan:2008zzk}.

\section{Data sets and simulated event samples}
\label{sec:data}

The collision data used in this analysis is collected with an inclusive single-muon (electron) trigger.
The events in the dimuon channel are triggered by the presence of at least one
muon candidate with $\pt > 20\GeV$ and $\abs{\eta} < 2.4$.
The events for the dielectron channel are triggered by an electron with $\pt >  23\GeV$ and $\abs{\eta} < 2.5$, satisfying loose identification and isolation criteria.

Various MC samples are used to simulate the DY signal and background processes.
The \MGvATNLO v2.2.2~\cite{Alwall:2014hca} event generator is used to simulate the signal and \PW~production in association with one or more jets ($\PW$+jets)
events at NLO accuracy in the QCD coupling constant, \alpS. Pair production of top quarks (\ttbar) and single top quark production in association with a \PW~boson ($\PQt\PW$~and $\PAQt\PW$) are generated at NLO using \POWHEG v2.0 and v1.0 respectively~\cite{POWHEG1,POWHEG2,POWHEG3,POWHEG-ttbar, POWHEG-tW}.
Diboson processes ($\PW\PW$, $\PW\cPZ$, and $\cPZ\cPZ$) are simulated at leading order with \PYTHIA v8.212~\cite{Sjostrand:2014zea}.
All samples are generated with the PDF package of NNPDF3.0~\cite{NNPDF,NNPDF30}.
The \PYTHIA generator with the underlying event tune CUETP8M1~\cite{Khachatryan:2015pea} is used
for the showering and hadronisation in all samples.
For simulations at NLO, jets from matrix element calculations and parton showering are merged using the FxFx prescription~\cite{FxFx}.
The total production rate for each process is normalised using the most accurate theoretical cross section value available.
The DY process is normalised to the predicted cross section calculated using the \MGvATNLO.
The \ttbar rate is normalised to the predicted cross section using a calculation performed with NNLO+NNLL (next-to-next-to-leading logarithm) accuracy~\cite{Czakon:2011xx}. The normalisations for the single top quark and diboson samples use cross section values available at NLO accuracy~\cite{Campbell:2011bn, Melia:2011tj}.

In all the MC samples, the detector response is simulated using a detailed description of the CMS
detector based on the \GEANTfour~\cite{Agostinelli:2002hh} package.
The simulated events are reconstructed using the same software as the real data.
Minimum bias events are superimposed on the simulated physics processes
to emulate the effects of multiple interactions per bunch crossing (pileup);
typically an average number of 11.
All MC samples are reweighted to provide the same pileup distribution observed in the data.

\section{Event selection}
\label{sec:sel}

Each reconstructed offline muon candidate is required to meet identification criteria that are based on the number of hits found in the tracker,
the response of the muon detectors, and a set of criteria based on the matching between the muon track parameters as measured by the inner tracker and muon detectors. Furthermore, the two muon candidates are required to share a well-defined common vertex. To reject cosmic ray muons that can appear to be back-to-back muon pairs, the transverse impact parameter with respect to the centre of the interaction region is required to be small and the opening angle between the two candidates should be differ from $\pi$ by more than 5 mrad. In order to suppress nonprompt muon candidates from heavy flavour decays, muons are required to be well-isolated within a cone of size $\Delta R = 0.3$, where $\Delta R = \sqrt{\smash[b]{(\Delta\eta)^2+(\Delta\phi)^2}}$.
More details on the muon reconstruction, identification and isolation criteria used in this analysis can be found in Refs.~\cite{CMS-PAPER-MUO-16-001}.

The offline reconstructed electron candidates are required to pass identification criteria that are based on electromagnetic shower shape variables. Electron candidates originating from photon conversions are suppressed by requiring that they have at most two missing inner tracker hits and that they are not consistent with being part of a conversion pair. The isolation of an electron candidate is defined by measuring the sum of energy deposits associated with the photons as well as with the charged and neutral hadrons reconstructed by the particle-flow (PF) algorithm~\cite{PFT2017} with the same cone size as for the muons.
The electron selection is based on the ratio of the PF isolation to the \pt of the electron candidate.
More details about electron reconstruction and identification criteria used in this analysis are given in Refs.~\cite{bib:EGM11001,CMS:EGM-13-001}.

For the offline analysis the leading muon (electron) candidate in the event is required to have $\pt > 22~(30)\GeV$ and the subleading candidate $\pt > 10\GeV$.
All muon candidates are required to satisfy $\abs{\eta} < 2.4$, and all electron candidates should satisfy $\abs{\eta} < 2.5$, while excluding the barrel-endcap transition region of the ECAL ($1.44 < \abs{\eta} < 1.57$).
The two muon candidates are required to have opposite charges, and the pair with the best fit
 of the dimuon vertex is selected if there is more than one candidate pair in the same event.
No opposite-charge and same-vertex requirements are applied in the dielectron channel
to avoid a selection inefficiency. At least one of the two candidate leptons, selected in the event,
is required to match the object that triggered the events.

The reconstructed dilepton invariant mass distribution can be affected by an imperfect measurement of the momentum and energy of the lepton candidates.
Momentum scale corrections for the muons are applied using the procedures described in Refs.~\cite{CMSDY3, bib:momcor}.
The electron energy deposits, as measured in ECAL, are subject to a set of
corrections involving information from both the ECAL and tracker~\cite{bib:EGM11001,CMS:EGM-13-001}.

The measurements are performed in 43 bins of dilepton invariant mass.
The binning at $m < 600\GeV$ is identical to that used in the earlier measurement~\cite{CMSDY3},
although the highest bin is extended to 3000\GeV.
The highest mass events observed in the data set are about 2.3\TeV in both channels.

\section{Background estimation}
\label{sec:bg}

The background composition varies across the dilepton invariant mass range, with the dominant backgrounds over the entire mass range being \ttbar, $\PQt\PW$, and $\PAQt\PW$ production, except for the region below the \cPZ~boson peak, where DY production of $\tau^+\tau^-$ pairs and their subsequent decays to electron and muon pairs constitute a significant background.
In the dimuon channel, the QCD multijet background is relatively large at low mass, below about 60\GeV.
However, in the dielectron channel, it contributes significantly in the high mass region as well.

The main backgrounds are estimated from data using control samples to reduce the systematic uncertainty due to the imperfect knowledge of the theoretical cross sections for the SM processes. For \ttbar, $\PQt\PW$, $\PAQt\PW$, $\tau^+\tau^-$, and $\PW\PW$ production, the contributions are estimated from a large data sample containing electron-muon pairs of opposite charge; the identification criteria used to select this sample are the same used to select the signal samples (described in the previous section).
The rates are twice as large as those for dimuon or dielectron events; however, the contributions of the same-flavour final states are scaled after correcting for the relevant detector acceptance and efficiency.

The QCD multijet and $\PW$+jets processes contribute to the background mainly when one or more jets are misidentified as electron or (less likely) as muon candidates;
the contributions are estimated using the ``misidentification rate'' method
described in Refs.~\cite{EXO-12-061,CMSDY2}. The misidentification rate is measured as a function of muon and electron \pt
in the barrel and endcap regions separately.
Subsequently, the misidentification rate is applied to events with loosely isolated leptons, after subtracting the contributions of genuine dilepton events.
The contributions from genuine lepton candidates from DY and \ttbar production are subtracted using a lepton distribution fit obtained from the simulation. The misidentification rate in the dimuon channel is defined as
the ratio of the number of isolated muon candidates passing the muon identification criteria and the number of muon candidates passing  the muon identification with no isolation requirement.
For this measurement a data sample selected with the single-muon trigger is used. The definition of the misidentification rate in the dielectron channel is slightly different and is calculated by measuring the probability of a jet passing the final electron selection criteria described above. For this measurement, a large unbiased event sample has been used, selected with a combination of single-photon triggers rather than electron triggers. The fraction of misidentified lepton backgrounds is generally less than 1\% across the entire mass range in the dimuon channel and up to 3\% (5\%) below (above) the \cPZ~boson peak in the dielectron channel.

The contribution from $\PW\cPZ$ and $\cPZ\cPZ$ processes are evaluated using simulation. The photon-initiated (PI) production of same-flavour lepton pairs is estimated with the \FEWZ 3.1 package~\cite{bib:FEWZ,bib:FEWZ1}, using the LUXqed photon PDF~\cite{bib:LUXqed}. Events generated with \PYTHIA are used in the MRST2004qed PDF set as a cross-check~\cite{Bourilkov:2016qum}. The contributions from PI production are negligible, except in the high mass region~\cite{Bourilkov:2016oet}.

\begin{figure}[htpb]
{\centering
\includegraphics[width=0.75\textwidth]{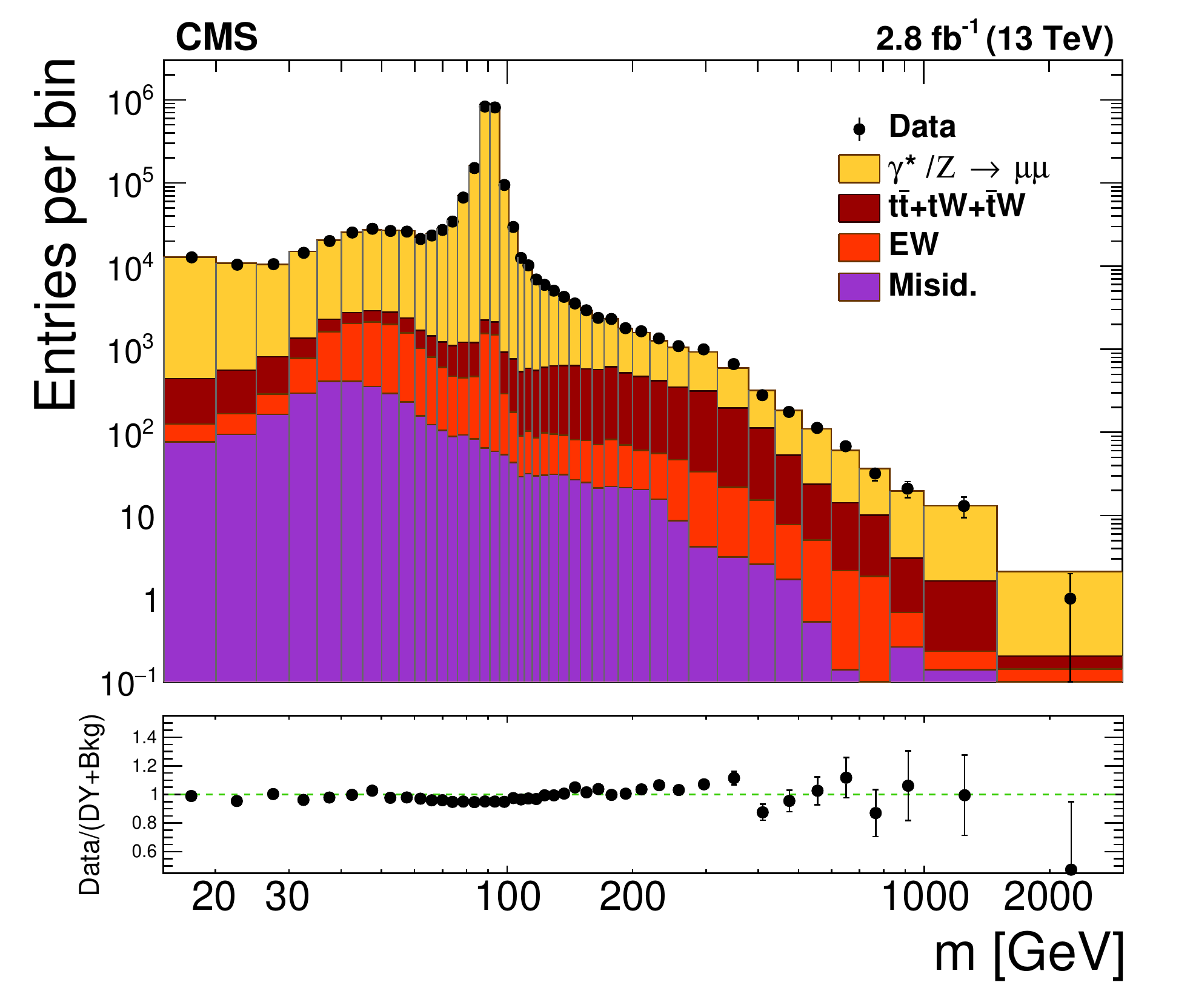}
\includegraphics[width=0.75\textwidth]{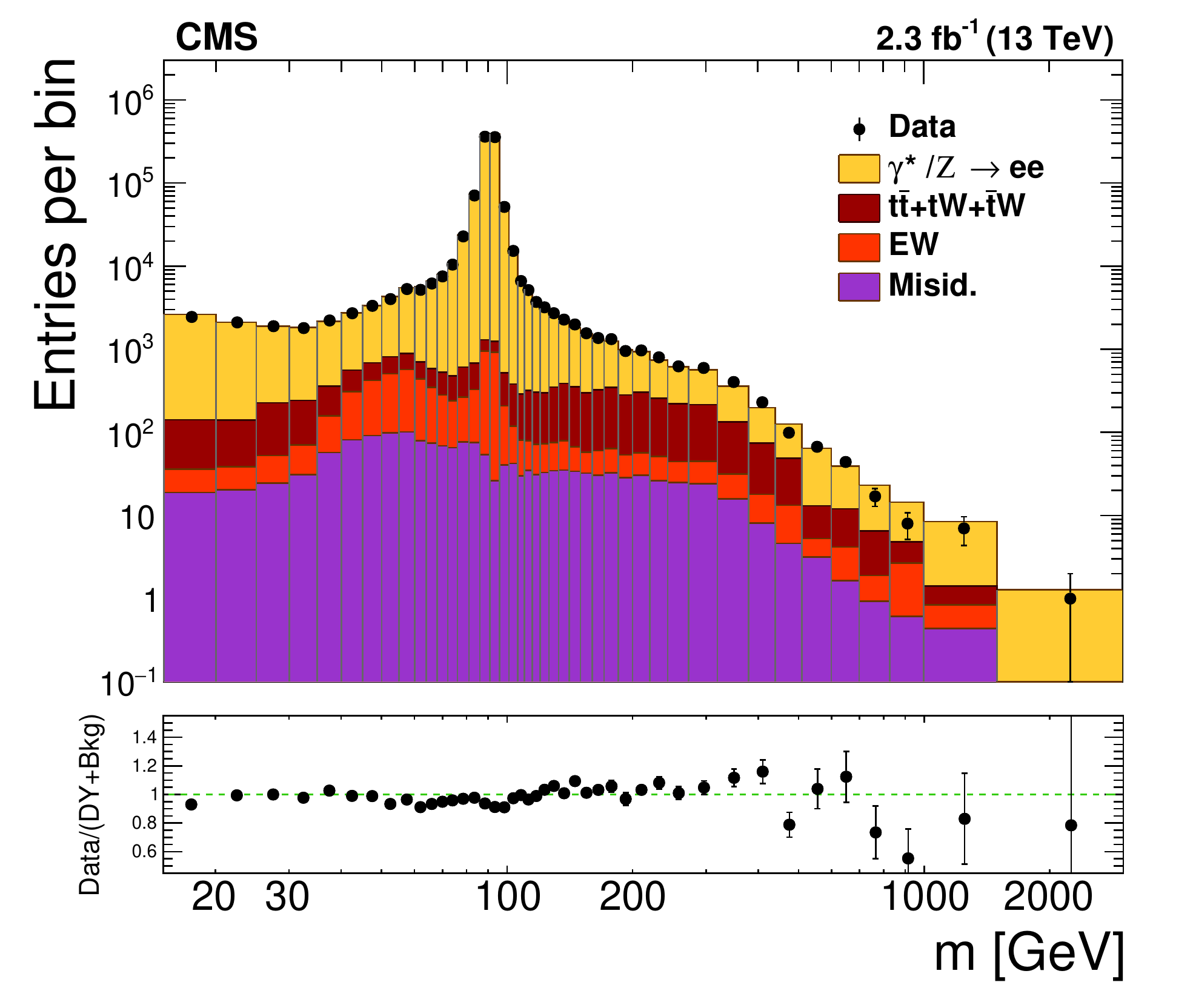}
\caption{
\label{fig:invMass}
The observed dimuon (top) and dielectron (bottom) invariant mass spectra within the detector acceptance.
The ``EW'' label indicates the contributions from the $\mathrm{DY}$ production of $\tau^+\tau^-$, $\PW\PW$, $\PW\cPZ$, and $\cPZ\cPZ$ processes.
The ``Misid.'' label corresponds \PW+jets and QCD multijet backgrounds.
Each MC process is normalised using the most accurate theoretical cross section value available.
The error bars on the data points represent the statistical uncertainty only.
}
}
\end{figure}

Figure~\ref{fig:invMass} presents the dilepton invariant mass distributions in dimuon and dielectron channels.
The cumulative yields from the signal and the background discussed above are superimposed on the data.

\section{Corrections}
\label{sec:corr}

To compare the measured distributions with the theoretical predictions, various experimental corrections need to be applied after subtracting the total expected background from the observed number of events in each mass bin. The correction for the detector resolution effects is implemented using an unfolding technique, similar to that used in the previous measurement~\cite{CMSDY3}. The event acceptance and selection efficiency are estimated using simulation and are used to correct the data. Any difference in the selection efficiency between the data and simulation is corrected for using a scale factor.
Finally, the correction required to account for the effects of FSR on the invariant mass of the lepton pair is applied, using so-called ``dressed leptons'' and further unfolding. The details of the correction procedures are discussed below.

\subsection{Detector resolution effects}
\label{sec:detrel}

The detector resolution occasionally leads to migration of events from mass bin $k$ of the generator level distribution to another bin $i$ of the reconstructed distribution. This effect is corrected by unfolding using the iterative D'Agostini method~\cite{dagostini} with a response matrix.
The iteration is terminated when the difference between the result and the previous one is less then 0.1\% in the entire mass bin. The validity of the unfolding method is tested on simulated events.

The elements of the detector response matrix, $T_{ik}$, are calculated using the DY simulation sample.
Each element contains the fraction of events that migrated from the $k$th bin of the generator level (the post--FSR level, described in Section~\ref{sec:fsr}) mass distribution to the $i$th bin of the reconstructed mass distribution, such that
\begin{equation}
N_{\text{obs},i} = \sum_{k} T_{ik}N_{\text{gen level},k}.
\end{equation}
The effect of the unfolding on the differential cross section is largest in the \cPZ~boson peak region because of the narrow width, where the effect on the signal yield is observed as large as $30$ and $40\%$ in the dimuon and dielectron channels, respectively.

\subsection{Acceptance and efficiency}
\label{sec:acceff}

\begin{figure}[htpb]
{\centering
\includegraphics[width=0.75\textwidth]{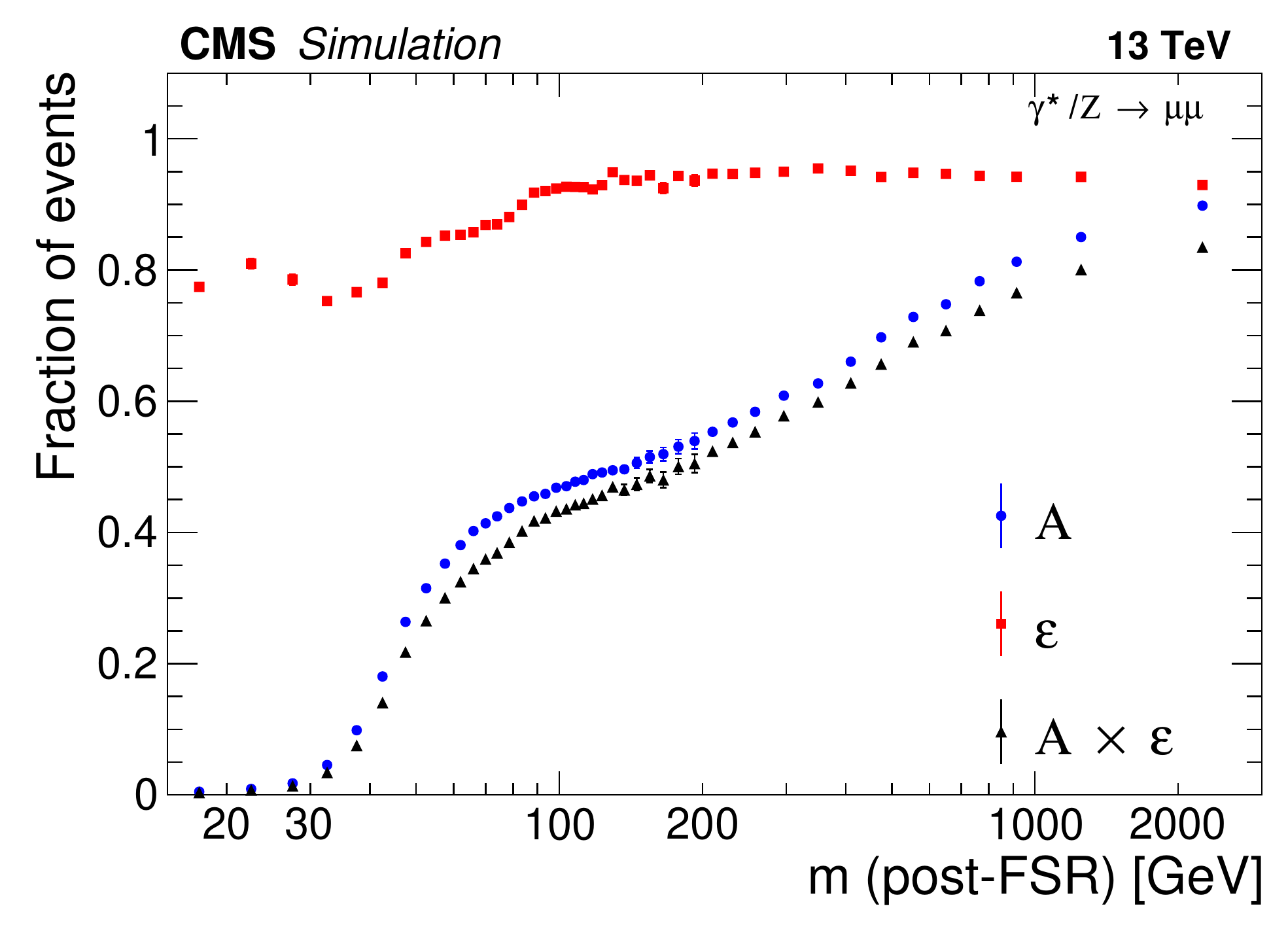}
\includegraphics[width=0.75\textwidth]{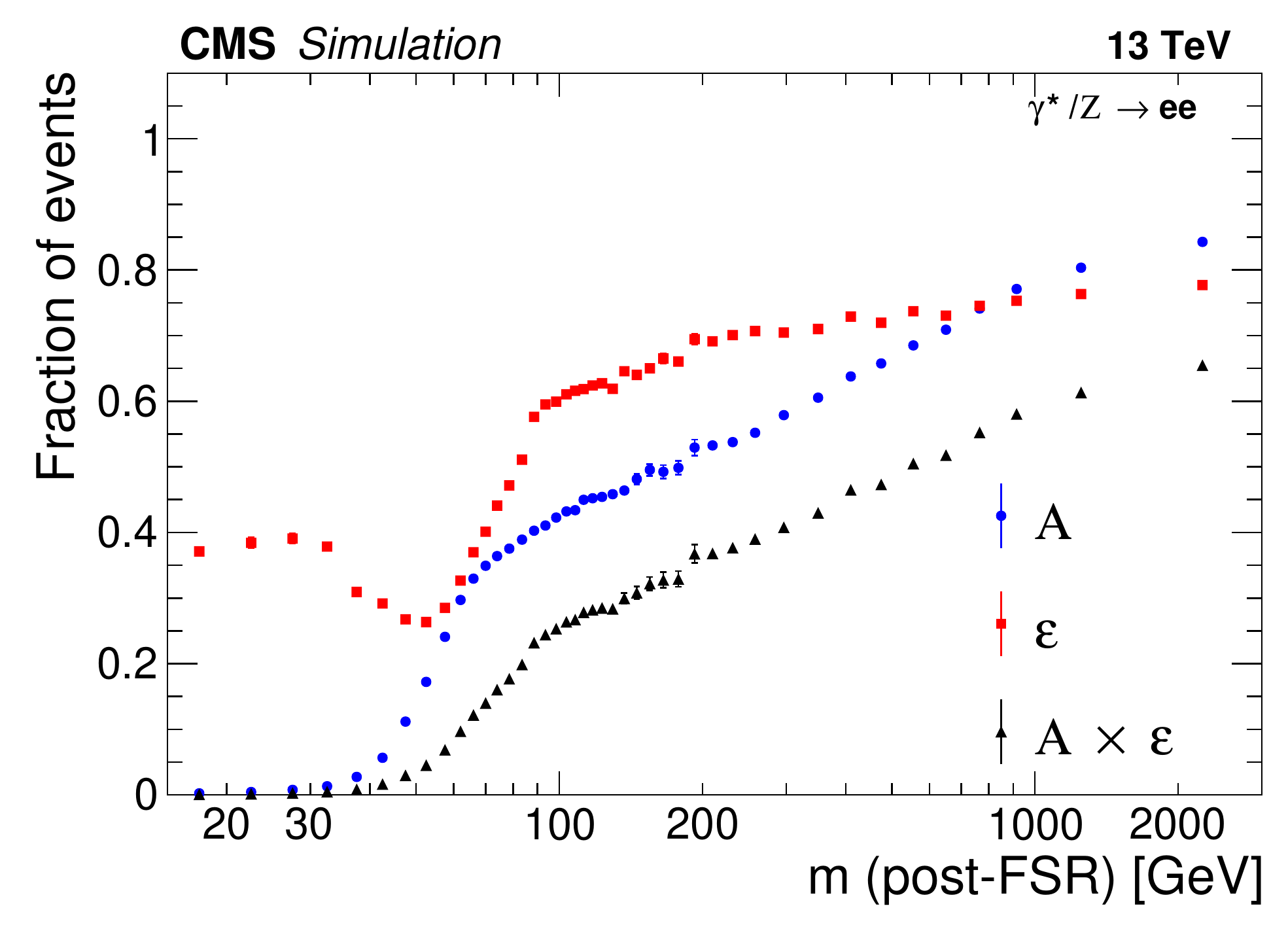}
\caption{
  \label{fig:1Dacceff}
  The signal acceptance (A), efficiency ($\varepsilon$) and their product
  for each invariant mass bin in the dimuon (top) and dielectron (bottom) channels,
  calculated from simulation.
  The error bars on the data points represent the statistical uncertainty only.
}
}
\end{figure}

The acceptance is defined as the fraction of simulated signal events with both leptons passing the nominal
\pt and $\eta$ requirements of the analysis mentioned in Section~\ref{sec:sel}, with respect to the full phase space.
The efficiency is defined as the fraction of simulated signal events that lie within the acceptance and that satisfy all the event selection criteria. The following equation defines the acceptance and efficiency for a given mass bin:
\begin{equation}
A \, \varepsilon = \frac{N_{\mathrm{A}}}{N_{\text{gen}}} \, \frac{N_{\varepsilon}}{N_{\mathrm{A}}},
\end{equation}
where $N_{\text{gen}}$ is the total number of generated signal events,
$N_{\mathrm{A}}$ is the number of events passing the acceptance criteria,
and $N_{\varepsilon}$ is the number of reconstructed events passing the full event selection.
Figure~\ref{fig:1Dacceff} shows the variation of the acceptance and efficiency as functions of the dilepton invariant mass in the dimuon and dielectron channels.
The acceptance correction is not applied to the fiducial cross section, but it is applied to the result in the full phase space, as described in Section~\ref{sec:result}.

\subsection{Efficiency correction}
\label{sec:effcorr}

To correct for the difference in efficiencies between
data and simulation for lepton reconstruction, identification, isolation, and trigger, scale factors
are determined from the data using the ``tag-and-probe'' method~\cite{bib:CMS_WZ}.
Events are selected that contain a dilepton pair near the resonance region where the background is very small.
One lepton is required to satisfy the tight selection and the other is used as the probe lepton.
The efficiency is then determined from the number of probe leptons that also pass the required selection criteria.
The measured efficiency using the tag-and-probe method is parametrised as a function of lepton \pt and $\eta$
and is then factorised into the reconstruction, identification, and isolation related components.
The overall efficiency is then given by
\begin{equation}
\varepsilon_{\mathrm{event}} = \varepsilon_{\ell_1} \, \varepsilon_{\ell_2} \, \varepsilon_{\text{event, trig}},
\end{equation}
where $\varepsilon_{\ell _{1,2}}$ are the single-lepton efficiencies of the individual leptons and $\varepsilon_{\text{event, trig}}$ is the trigger efficiency for the event~\cite{CMSDY2, CMSDY3}.

The scale factor between data and simulation is determined
by the ratio $\varepsilon_{\text{data}}(\text{event}) / \varepsilon_{\mathrm{MC}}(\text{event})$.
The scale factors are measured to be 0.92--0.97 (0.95--0.97) in the dimuon (dielectron) channel;
the values are dependent on the data-taking period.
The scale factors are then applied to the simulation to correct for the observed differences.

\subsection{Final-state photon radiation effects}
\label{sec:fsr}

The FSR from a lepton shifts the measured invariant mass of the dilepton pair to lower values, which significantly affects
the distribution below the \cPZ~boson peak, especially in the dielectron channel. The definition of ``dressed lepton'' accounts for the required  correction for a given flavour so that the results from the individual channels can be combined subsequently for comparison with the theoretical predictions. The four-momentum of the ``dressed lepton'' is defined as
\begin{equation}
{p}^{\text{dressed}}_{\ell} = {p}^{\text{post--FSR}}_{\ell} + \Sigma{p}_{\gamma},
\end{equation}
where the four-momenta of all the simulated photons originating from the leptons are summed within a cone of $\Delta R < 0.1$ around the candidate lepton, where $\Delta R$ is the separation of the photon from the lepton in $\eta$-$\phi$ space. This enables to correct the lepton momentum to compensate missing momentum due to the FSR.

The FSR correction is estimated separately using an unfolding technique from a signal MC sample where the FSR is simulated by \PYTHIA. The response matrix is produced using the information about dressed and post--FSR leptons in the simulation.
The correction to the cross section, defined in terms of dressed leptons using data corresponding to a post--FSR lepton, is in the range of 0.78--1.09 (0.58--1.27) in the dimuon (dielectron) channel.
The FSR correction is not applied to the fiducial cross section, but it is applied to the result in the full phase
space described in Section~\ref{sec:result}.

\section{Systematic uncertainties}
\label{sec:syst}

Systematic uncertainties are the dominant source of the total uncertainty in this measurement for $m < 400$\GeV.

In the dimuon channel, the efficiency scale factor (which includes muon reconstruction, identification, isolation, and trigger selection) is the most significant component of the systematic uncertainty below the \cPZ~boson peak. A variety of possible contributions to the efficiency scale factor are listed:
\begin{itemize}
\item statistical uncertainty associated with the tag-and-probe procedure;
\item binning in muon \pt and $\eta$;
\item shape hypotheses for the signal and the background in the fit model;
\item other minor sources, such as the number of mass bins chosen, the selected mass range, and the kinematic selection criteria.
\end{itemize}

Uncertainties for all the sources are evaluated separately and are combined in quadrature.

Detector resolution effects, including the muon momentum scale correction, are major sources of systematic uncertainty in the dimuon channel.
Both data and simulation are smeared by varying the muon momentum scale within its uncertainty.
The difference between the cross section values obtained with and without the smearing is assigned as a systematic uncertainty in each mass bin.  In addition, contributions to the uncertainty relating to the unfolding of the detector resolution are:

\begin{itemize}
\item the statistical uncertainty in the response matrix due to the finite size of the MC sample;
\item the systematic uncertainty in the response matrix arising from the differences in the MC modeling, as determined by comparing
two different MC generators: \POWHEG and \MGvATNLO.
\end{itemize}

In the dielectron channel, the dominant systematic uncertainty below the \cPZ~boson peak is the efficiency scale factor, as in the dimuon channel.
Though the method to calculate this uncertainty is therefore similar, the possible sources are slightly different from those in the dimuon channel:

\begin{itemize}
\item statistical uncertainty associated with the tag-and-probe procedure;
\item shape hypotheses for the signal and the background in the fit model;
\item difference between the NLO and LO MC samples;
\item different selection applied to the tag electron.
\end{itemize}

Detector resolution effects, including the electron energy scale and smearing corrections, are another significant systematic uncertainty in the dielectron channel. This uncertainty is determined with a method similar to that used in the dimuon channel, where the simulation is varied by the electron energy scale correction and smeared with the uncertainties. The difference between the central value and the smeared result for the cross section is assigned as the uncertainty. In addition, two other sources of systematic uncertainty in the detector resolution unfolding are considered for the dielectron channel: the statistical uncertainty in the response matrix and the difference between MC models. The uncertainty in the MC modeling is derived by comparing the results based on \MGvATNLO with those based on events reweighted with \FEWZ package.

In both channels, the statistical uncertainty in the data sample, used for the background estimation, is one of the sources of systematic uncertainty; it is the dominant component  in the higher mass region, above 200\GeV.

The difference between the background contributions estimated using control samples in data and the predictions from the MC simulation is one of the systematic uncertainty sources.
In the cases where the background rates are estimated from MC simulation,
the corresponding uncertainty in the predicted cross section is included as a systematic uncertainty
although the contribution is negligible compared with other uncertainty components in the entire mass range.

The systematic uncertainty related to the acceptance is dominated by the theoretical component, which originates mainly from imperfect knowledge of nonperturbative effects, such as the PDFs.
The PDF uncertainties are estimated using the NNLO version of \FEWZ package
and the FSR correction is applied in the calculation by using the dressed-lepton definition.

The difference between \MGvATNLO and \FEWZ in the prediction of the DY differential cross section
is assigned as an additional uncertainty in the acceptance.
The uncertainty in the cross section due to the variation of the strong coupling constant, \alpS,
is estimated by varying the input value for the NNLO PDF used along with \FEWZ.

The model-dependent FSR simulation is another source of uncertainty and it is evaluated by comparing the results from the \PYTHIA and \PHOTOS 3.56~\cite{PHOTOS3} generators.
The difference in the cross section, after the FSR correction using dressed leptons, between the unfolding procedures of the \PYTHIA and \PHOTOS generators is assigned as a systematic uncertainty.

\begin{figure} [htpb]
{\centering
\includegraphics[width=0.77\textwidth]{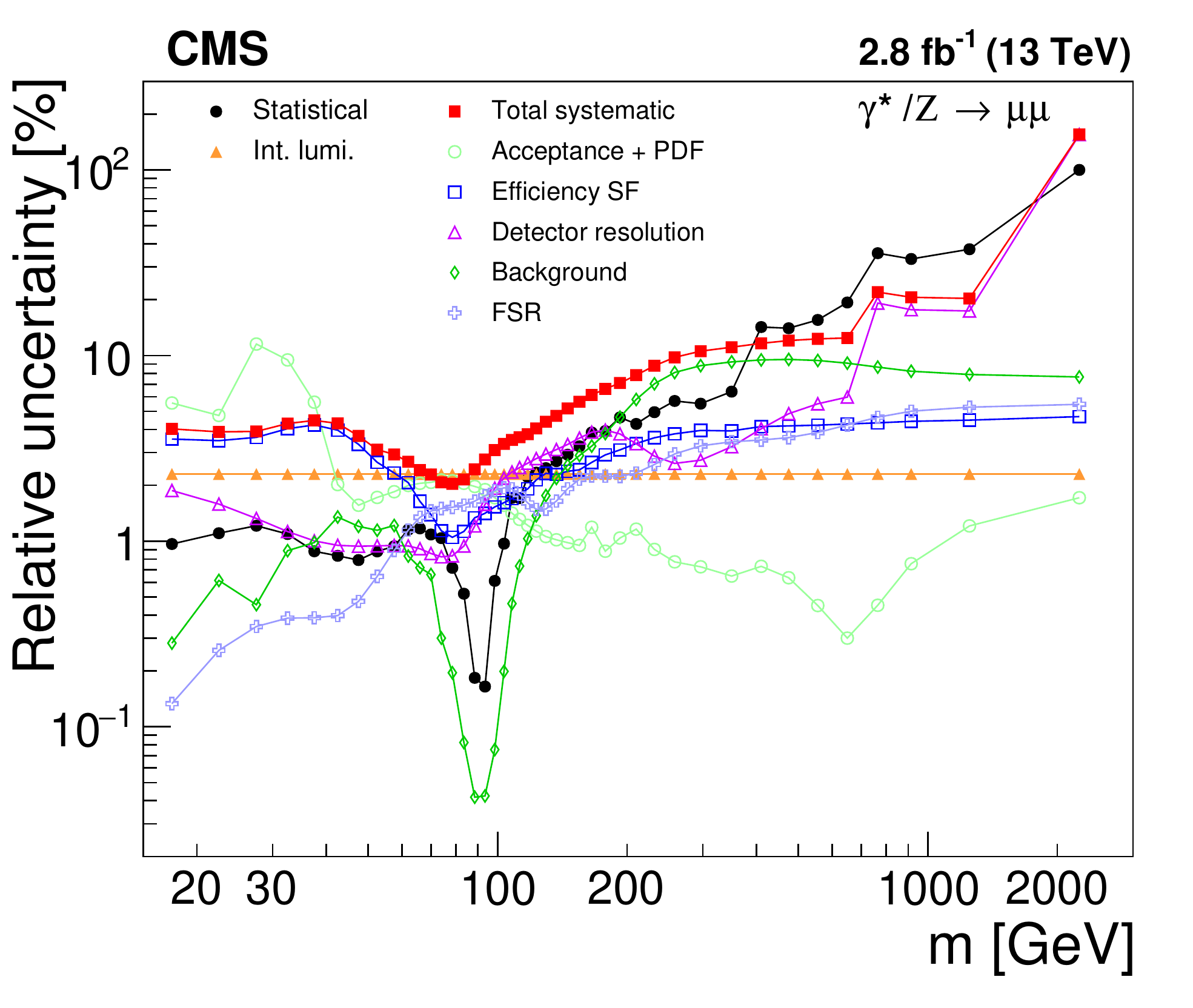}
\includegraphics[width=0.77\textwidth]{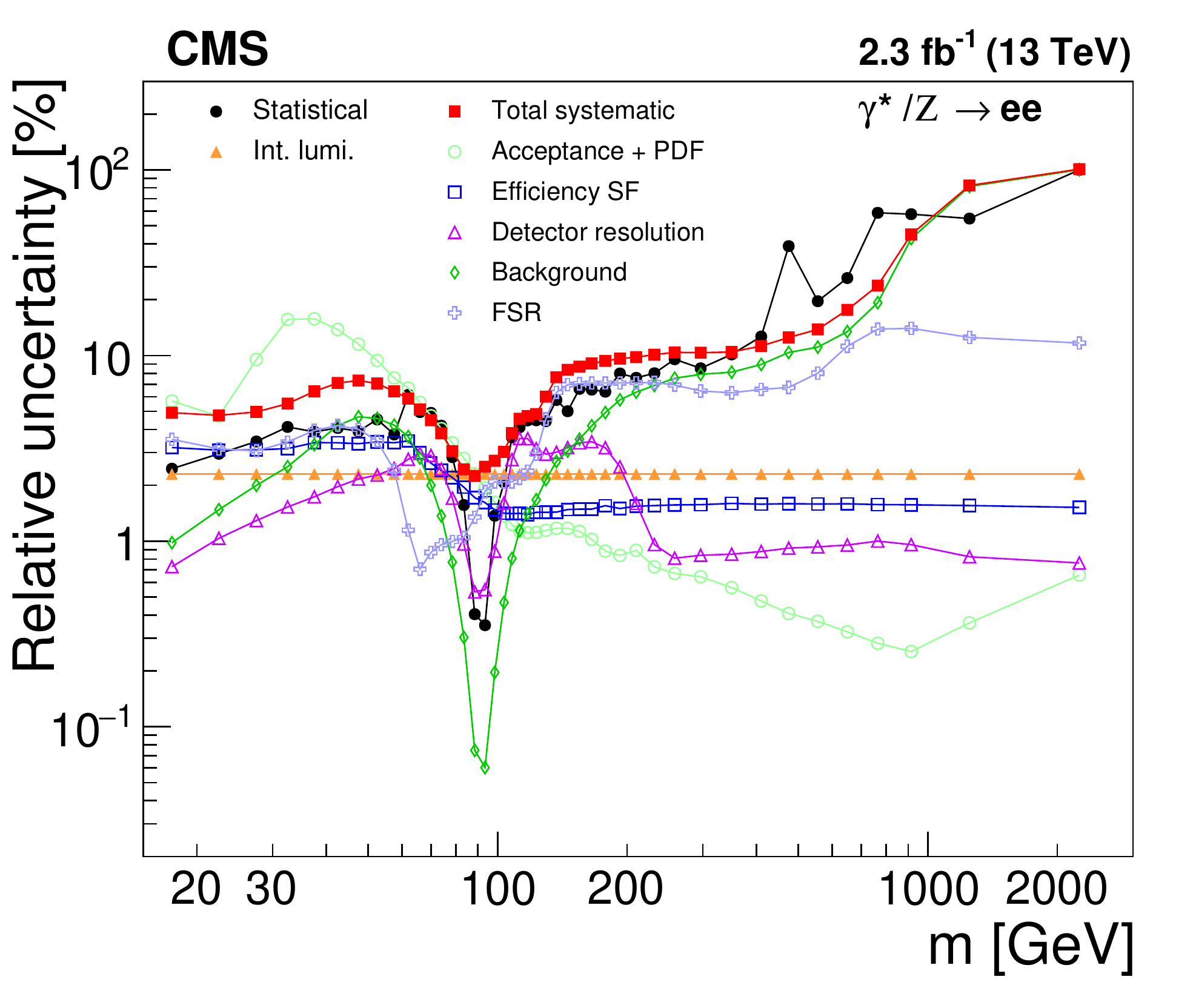}
\caption{
\label{fig:syst-muons}
  Summary of the systematic uncertainties on the differential cross section measurement $\rd\sigma / \rd{m}$ [pb/\GeVns{}]  in the dimuon (upper) and dielectron (lower) channels.
The ``Total systematic'' is a quadratic sum of all systematic uncertainty sources except for the ``Acceptance + PDF''.
}
}
\end{figure}

The uncertainty on the integrated luminosity measurement, based on pixel cluster counting in the silicon pixel detector, is 2.3\%~\cite{CMS-PAS-LUM-15-001}.
Figure~\ref{fig:syst-muons} shows the estimated systematic uncertainties obtained in each mass bin;
these are also summarised in Table~\ref{tab:syst-muons} (\ref{tab:syst-electron}) for the dimuon (dielectron) channel. The uncertainty due to the acceptance and the PDFs originates from incomplete theoretical knowledge and is categorised separately from the other components, which are experimental in nature and are labelled ``Total''.

\begin{table} [htpb]
\centering
\topcaption{Summary of the systematic uncertainties (\%) for the $\rd\sigma / \rd{m}$ (pb/\GeVns{}) measurement in the dimuon channel. The column labelled ``Total'' corresponds to the quadratic sum of all the experimental sources, except for that Acceptance+PDF.
}
\label{tab:syst-muons}
\begin{tabular}{cccccccc}
\hline
$m$ & Eff. & Det. resol. & Bkgr. est. & FSR & Total & Acceptance+PDF \\
  (\GeVns{}) & (\%) & (\%) & (\%) & (\%)& (\%) & (\%) \\
\hline
15--20 & $  3.5  $ & $  1.9  $ & $  0.28  $ & $  0.13  $ & $  4.0  $ & $  5.6  $\\
20--25 & $  3.5  $ & $  1.6  $ & $  0.61  $ & $  0.26  $ & $  3.9  $ & $  4.8  $\\
25--30 & $  3.6  $ & $  1.3  $ & $  0.45  $ & $  0.35  $ & $  3.9  $ & $  12  $\\
30--35 & $  4.0  $ & $  1.1  $ & $  0.89  $ & $  0.38  $ & $  4.3  $ & $  9.5  $\\
35--40 & $  4.2  $ & $  1.0  $ & $  0.98  $ & $  0.39  $ & $  4.5  $ & $  5.6  $\\
40--45 & $  4.0  $ & $  0.95  $ & $  1.3  $ & $  0.40  $ & $  4.3  $ & $  2.0  $\\
45--50 & $  3.3  $ & $  0.94  $ & $  1.2  $ & $  0.47  $ & $  3.7  $ & $  1.6  $\\
50--55 & $  2.7  $ & $  0.94  $ & $  1.1  $ & $  0.65  $ & $  3.1  $ & $  1.7  $\\
55--60 & $  2.3  $ & $  0.95  $ & $  1.2  $ & $  0.89  $ & $  2.9  $ & $  1.8  $\\
60--64 & $  2.1  $ & $  0.94  $ & $  0.83  $ & $  1.1  $ & $  2.7  $ & $  2.0  $\\
64--68 & $  1.6  $ & $  0.91  $ & $  0.72  $ & $  1.3  $ & $  2.4  $ & $  2.0  $\\
68--72 & $  1.4  $ & $  0.86  $ & $  0.66  $ & $  1.5  $ & $  2.3  $ & $  2.1  $\\
72--76 & $  1.1  $ & $  0.82  $ & $  0.30  $ & $  1.5  $ & $  2.1  $ & $  2.1  $\\
76--81 & $  1.0  $ & $  0.83  $ & $  0.20  $ & $  1.5  $ & $  2.0  $ & $  2.1  $\\
81--86 & $  1.1  $ & $  0.94  $ & $  0.082  $ & $  1.6  $ & $  2.1  $ & $  2.0  $\\
86--91 & $  1.3  $ & $  1.2  $ & $  0.042  $ & $  1.6  $ & $  2.4  $ & $  2.0  $\\
91--96 & $  1.4  $ & $  1.6  $ & $  0.042  $ & $  1.8  $ & $  2.8  $ & $  1.9  $\\
96--101 & $  1.5  $ & $  1.9  $ & $  0.075  $ & $  1.9  $ & $  3.1  $ & $  1.7  $\\
101--106 & $  1.6  $ & $  2.2  $ & $  0.20  $ & $  1.9  $ & $  3.3  $ & $  1.6  $\\
106--110 & $  1.7  $ & $  2.4  $ & $  0.46  $ & $  1.9  $ & $  3.5  $ & $  1.4  $\\
110--115 & $  1.8  $ & $  2.5  $ & $  0.73  $ & $  1.8  $ & $  3.6  $ & $  1.3  $\\
115--120 & $  1.9  $ & $  2.6  $ & $  1.0  $ & $  1.6  $ & $  3.8  $ & $  1.2  $\\
120--126 & $  2.1  $ & $  2.8  $ & $  1.4  $ & $  1.5  $ & $  4.0  $ & $  1.1  $\\
126--133 & $  2.3  $ & $  2.9  $ & $  1.8  $ & $  1.5  $ & $  4.4  $ & $  1.1  $\\
133--141 & $  2.3  $ & $  3.1  $ & $  2.2  $ & $  1.6  $ & $  4.7  $ & $  1.0  $\\
141--150 & $  2.4  $ & $  3.3  $ & $  2.6  $ & $  1.9  $ & $  5.2  $ & $  0.98  $\\
150--160 & $  2.4  $ & $  3.6  $ & $  2.9  $ & $  2.2  $ & $  5.6  $ & $  0.95  $\\
160--171 & $  2.7  $ & $  3.9  $ & $  3.3  $ & $  2.2  $ & $  6.1  $ & $  1.2  $\\
171--185 & $  2.9  $ & $  4.0  $ & $  3.8  $ & $  2.2  $ & $  6.6  $ & $  0.88  $\\
185--200 & $  3.1  $ & $  3.8  $ & $  4.7  $ & $  2.2  $ & $  7.1  $ & $  1.0  $\\
200--220 & $  3.4  $ & $  3.3  $ & $  5.8  $ & $  2.3  $ & $  7.8  $ & $  1.2  $\\
220--243 & $  3.6  $ & $  2.9  $ & $  7.0  $ & $  2.6  $ & $  8.8  $ & $  0.90  $\\
243--273 & $  3.8  $ & $  2.6  $ & $  8.1  $ & $  2.9  $ & $  9.8  $ & $  0.77  $\\
273--320 & $  3.9  $ & $  2.7  $ & $  8.8  $ & $  3.3  $ & $  11  $ & $  0.73  $\\
320--380 & $  3.9  $ & $  3.2  $ & $  9.2  $ & $  3.4  $ & $  11  $ & $  0.65  $\\
380--440 & $  4.1  $ & $  4.0  $ & $  9.5  $ & $  3.5  $ & $  12  $ & $  0.73  $\\
440--510 & $  4.2  $ & $  4.9  $ & $  9.5  $ & $  3.6  $ & $  12  $ & $  0.64  $\\
510--600 & $  4.2  $ & $  5.5  $ & $  9.4  $ & $  3.8  $ & $  12  $ & $  0.45  $\\
600--700 & $  4.3  $ & $  6.0  $ & $  9.1  $ & $  4.2  $ & $  12  $ & $  0.30  $\\
700--830 & $  4.3  $ & $  19  $ & $  8.7  $ & $  4.7  $ & $  22  $ & $  0.45  $\\
830--1000 & $  4.4  $ & $  18  $ & $  8.2  $ & $  5.0  $ & $  21  $ & $  0.76  $\\
1000--1500 & $  4.5  $ & $  17  $ & $  7.9  $ & $  5.3  $ & $  20  $ & $  1.2  $\\
1500--3000 & $  4.7  $ & $  150  $ & $  7.7  $ & $  5.5  $ & $  160  $ & $  1.7  $\\
\hline
\end{tabular}
\end{table}

\begin{table} [htpb]
\centering
\topcaption{Summary of the systematic uncertainties (\%) for the $\rd\sigma / \rd{m}$ (pb/\GeVns{}) measurement in the dielectron channel. The column labelled ``Total'' corresponds to the quadratic sum of all the experimental sources, except for that Acceptance+PDF.
}
\label{tab:syst-electron}
\begin{tabular}{cccccccc}
\hline
$m$ & Eff. & Det. resol. & Bkgr. est. & FSR & Total & Acceptance+PDF \\
  (\GeVns{}) & (\%) & (\%) & (\%) & (\%) & (\%) & (\%) \\
\hline
15--20 & $  3.2  $ & $  0.73  $ & $  0.98  $ & $  3.5  $ & $  4.9  $ & $  5.7  $\\
20--25 & $  3.1  $ & $  1.0  $ & $  1.5  $ & $  3.1  $ & $  4.8  $ & $  4.7  $\\
25--30 & $  3.1  $ & $  1.3  $ & $  2.0  $ & $  3.1  $ & $  5.0  $ & $  9.5  $\\
30--35 & $  3.2  $ & $  1.5  $ & $  2.5  $ & $  3.4  $ & $  5.5  $ & $  16  $\\
35--40 & $  3.4  $ & $  1.7  $ & $  3.3  $ & $  3.9  $ & $  6.4  $ & $  16  $\\
40--45 & $  3.4  $ & $  2.0  $ & $  4.2  $ & $  4.2  $ & $  7.1  $ & $  14  $\\
45--50 & $  3.4  $ & $  2.2  $ & $  4.7  $ & $  4.0  $ & $  7.3  $ & $  12  $\\
50--55 & $  3.4  $ & $  2.3  $ & $  4.6  $ & $  3.5  $ & $  7.1  $ & $  9.4  $\\
55--60 & $  3.4  $ & $  2.5  $ & $  4.2  $ & $  2.4  $ & $  6.4  $ & $  7.6  $\\
60--64 & $  3.5  $ & $  2.8  $ & $  3.7  $ & $  1.1  $ & $  5.9  $ & $  6.7  $\\
64--68 & $  3.0  $ & $  3.0  $ & $  2.8  $ & $  0.71  $ & $  5.1  $ & $  5.6  $\\
68--72 & $  2.6  $ & $  2.9  $ & $  2.0  $ & $  0.87  $ & $  4.5  $ & $  4.7  $\\
72--76 & $  2.4  $ & $  2.4  $ & $  1.4  $ & $  0.96  $ & $  3.8  $ & $  4.0  $\\
76--81 & $  2.2  $ & $  1.7  $ & $  0.77  $ & $  1.0  $ & $  3.1  $ & $  3.4  $\\
81--86 & $  2.0  $ & $  0.97  $ & $  0.30  $ & $  1.0  $ & $  2.4  $ & $  2.8  $\\
86--91 & $  1.7  $ & $  0.53  $ & $  0.075  $ & $  1.3  $ & $  2.2  $ & $  2.3  $\\
91--96 & $  1.6  $ & $  0.55  $ & $  0.060  $ & $  1.9  $ & $  2.5  $ & $  1.9  $\\
96--101 & $  1.5  $ & $  0.88  $ & $  0.20  $ & $  2.1  $ & $  2.7  $ & $  1.6  $\\
101--106 & $  1.4  $ & $  1.6  $ & $  0.47  $ & $  2.1  $ & $  3.0  $ & $  1.4  $\\
106--110 & $  1.4  $ & $  2.8  $ & $  0.81  $ & $  2.1  $ & $  3.8  $ & $  1.2  $\\
110--115 & $  1.4  $ & $  3.6  $ & $  1.1  $ & $  2.2  $ & $  4.6  $ & $  1.1  $\\
115--120 & $  1.4  $ & $  3.6  $ & $  1.4  $ & $  2.4  $ & $  4.7  $ & $  1.1  $\\
120--126 & $  1.4  $ & $  3.1  $ & $  1.7  $ & $  2.9  $ & $  4.8  $ & $  1.1  $\\
126--133 & $  1.4  $ & $  2.9  $ & $  2.1  $ & $  4.6  $ & $  6.0  $ & $  1.1  $\\
133--141 & $  1.4  $ & $  3.0  $ & $  2.7  $ & $  6.3  $ & $  7.6  $ & $  1.2  $\\
141--150 & $  1.5  $ & $  3.2  $ & $  3.1  $ & $  7.0  $ & $  8.4  $ & $  1.2  $\\
150--160 & $  1.5  $ & $  3.4  $ & $  3.5  $ & $  7.0  $ & $  8.7  $ & $  1.1  $\\
160--171 & $  1.5  $ & $  3.4  $ & $  4.2  $ & $  7.1  $ & $  9.1  $ & $  1.0  $\\
171--185 & $  1.6  $ & $  3.2  $ & $  4.9  $ & $  7.1  $ & $  9.4  $ & $  0.88  $\\
185--200 & $  1.5  $ & $  2.5  $ & $  5.8  $ & $  7.1  $ & $  9.6  $ & $  0.84  $\\
200--220 & $  1.5  $ & $  1.6  $ & $  6.3  $ & $  7.1  $ & $  9.8  $ & $  0.89  $\\
220--243 & $  1.6  $ & $  0.96  $ & $  6.9  $ & $  7.2  $ & $  10  $ & $  0.73  $\\
243--273 & $  1.6  $ & $  0.81  $ & $  7.5  $ & $  6.9  $ & $  10  $ & $  0.67  $\\
273--320 & $  1.6  $ & $  0.84  $ & $  7.9  $ & $  6.4  $ & $  10  $ & $  0.64  $\\
320--380 & $  1.6  $ & $  0.85  $ & $  8.1  $ & $  6.3  $ & $  10  $ & $  0.56  $\\
380--440 & $  1.6  $ & $  0.88  $ & $  9.0  $ & $  6.6  $ & $  11  $ & $  0.48  $\\
440--510 & $  1.6  $ & $  0.92  $ & $  10  $ & $  6.7  $ & $  13  $ & $  0.41  $\\
510--600 & $  1.6  $ & $  0.93  $ & $  11  $ & $  8.0  $ & $  14  $ & $  0.37  $\\
600--700 & $  1.6  $ & $  0.95  $ & $  13  $ & $  11  $ & $  18  $ & $  0.32  $\\
700--830 & $  1.6  $ & $  1.0  $ & $  19  $ & $  14  $ & $  24  $ & $  0.28  $\\
830--1000 & $  1.6  $ & $  0.96  $ & $  43  $ & $  14  $ & $  45  $ & $  0.25  $\\
1000--1500 & $  1.6  $ & $  0.82  $ & $  81  $ & $  13  $ & $  82  $ & $  0.36  $\\
1500--3000 & $  1.5  $ & $  0.76  $ & $  100  $ & $  12  $ & $  100  $ & $  0.66  $\\
\hline
\end{tabular}
\end{table}

\section{Results}
\label{sec:result}

\begin{figure}[htpb]
{\centering
\includegraphics[width=0.75\textwidth]{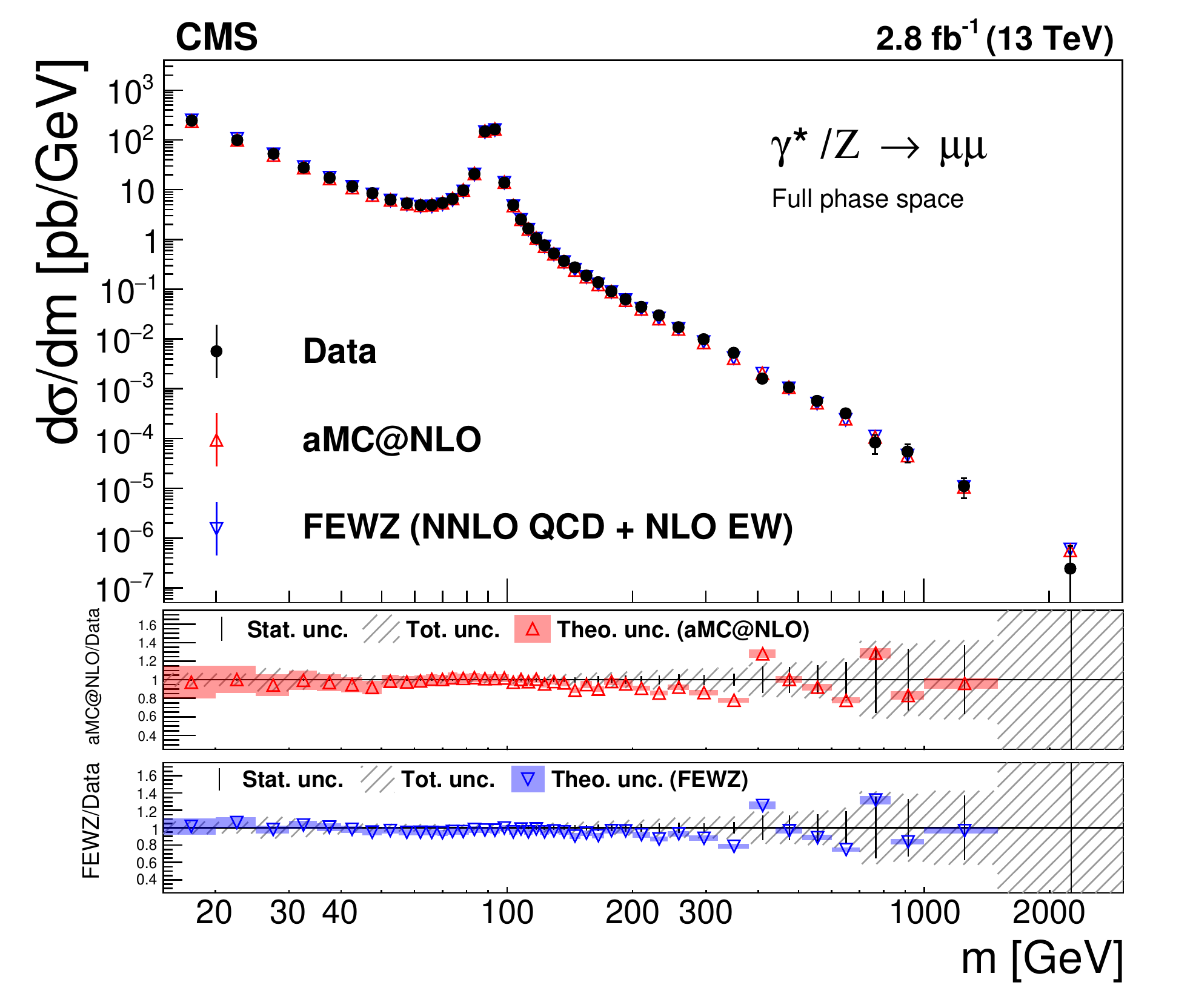}
\includegraphics[width=0.75\textwidth]{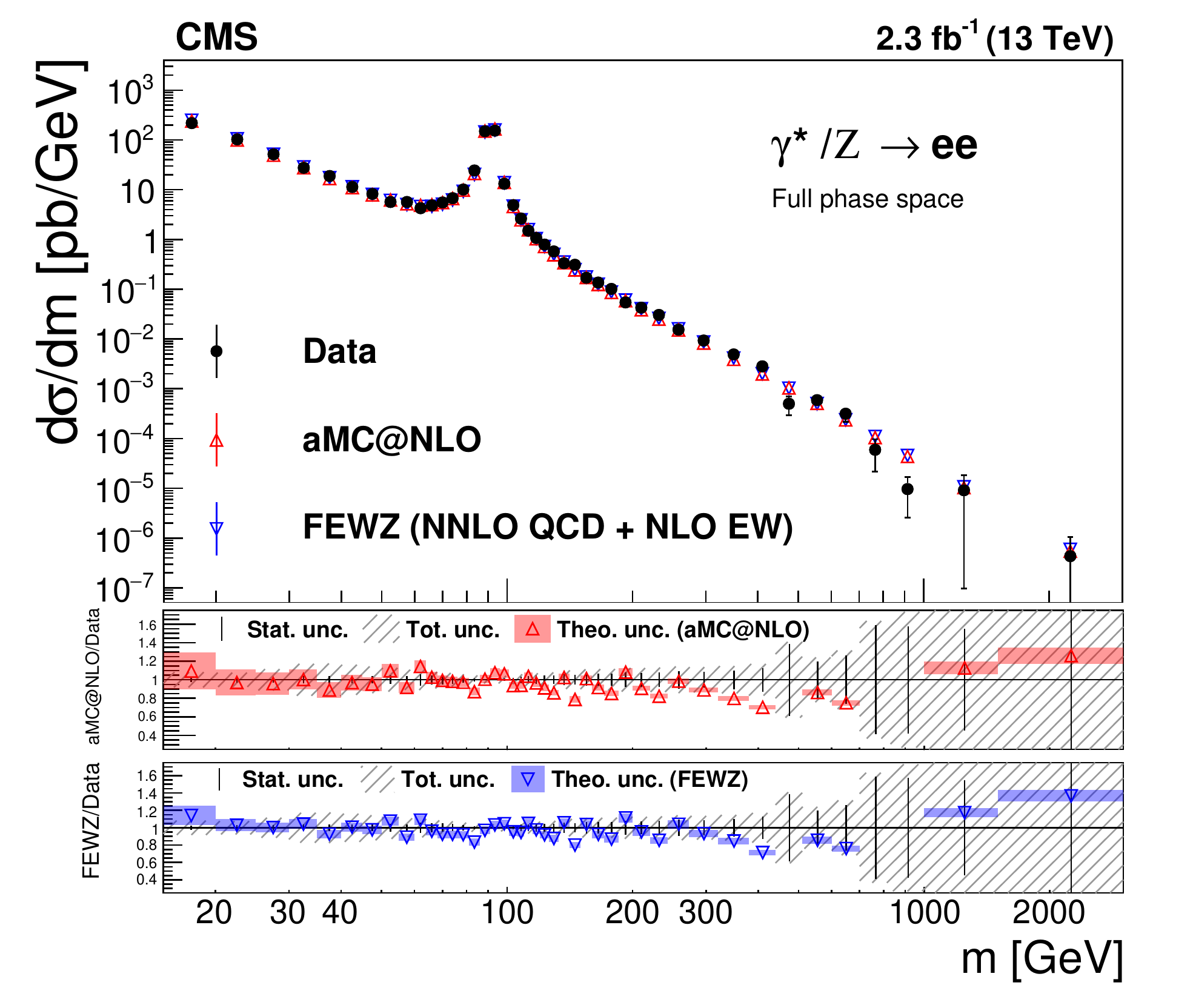}
\caption{
\label{fig:dsdM_fewz}
The differential cross section as a function of the dimuon (upper) and dielectron (lower) invariant mass, measured in the full phase space, with FSR correction applied. The spectra are compared with the NNLO theoretical prediction of \FEWZ (blue) and the NLO prediction of \MGvATNLO (red).  The NNPDF3.0 PDF set is used in both cases.  In the middle and lower panels, the coloured bands denote the theoretical uncertainty  and the hatched bands denote the total uncertainty, which is the combination of statistical, systematic, and integrated luminosity components.
}
}
\end{figure}

The differential cross section, in the full phase space, is estimated by extrapolation from the measurement within the fiducial region after the application of the set of corrections described in the previous sections.
The detector fiducial volume is defined by the \pt and $\eta$ requirements for the muons and electrons after FSR.
The results are presented in Fig.~\ref{fig:dsdM_fewz} as a function of
the dimuon and dielectron invariant mass covering the range 15 to 3000\GeV.
They are compared with the NNLO theoretical predictions from \FEWZ with NLO EW correction, as well as those from \MGvATNLO.
Both predictions are calculated using the NNPDF3.0 PDF set.
The ratios of the NLO and NNLO predictions to data are shown in the lower panels.
The measurements are in good agreement, within uncertainties, with both theoretical predictions.
Tables~\ref{tab:result-muons} and~\ref{tab:result-electrons} show the summary of the results for the
dimuon and dielectron channels, respectively.

\begin{table} [htpb]
\centering
\topcaption{Summary of the measured values of $\rd\sigma / \rd{m}$ (pb/\GeVns{}) in the dimuon channel with the statistical
($\delta_{\text{stat}}$), experimental ($\delta_{\text{exp}}$)  and theoretical ($\delta_{\text{theo}}$) uncertainties, respectively.
Here, $\delta_{\text{tot}}$ is the quadratic sum of the three components.
}
\label{tab:result-muons}
\begin{tabular}{ccccccc}
\hline
  $m(\GeVns{})$ & $\frac{\rd\sigma}{\rd{m}}$ (pb/\GeVns{}) & $\delta_{\text{stat}}$ & $\delta_{\text{exp}}$ & $\delta_{\text{theo}}$ & $\delta_{\text{tot}}$ \\
\hline
15--20 & $  2.5 \times 10^{2}  $ & $  2.4 \times 10^{0}  $ & $  1.1 \times 10^{1}  $ & $  1.4 \times 10^{1}  $ & $  1.8 \times 10^{1}  $\\
20--25 & $  9.9 \times 10^{1}  $ & $  1.1 \times 10^{0}  $ & $  4.5 \times 10^{0}  $ & $  4.7 \times 10^{0}  $ & $  6.6 \times 10^{0}  $\\
25--30 & $  5.3 \times 10^{1}  $ & $  6.4 \times 10^{-1}  $ & $  2.4 \times 10^{0}  $ & $  6.1 \times 10^{0}  $ & $  6.6 \times 10^{0}  $\\
30--35 & $  2.8 \times 10^{1}  $ & $  3.0 \times 10^{-1}  $ & $  1.4 \times 10^{0}  $ & $  2.6 \times 10^{0}  $ & $  3.0 \times 10^{0}  $\\
35--40 & $  1.7 \times 10^{1}  $ & $  1.5 \times 10^{-1}  $ & $  8.7 \times 10^{-1}  $ & $  9.7 \times 10^{-1}  $ & $  1.3 \times 10^{0}  $\\
40--45 & $  1.2 \times 10^{1}  $ & $  9.7 \times 10^{-2}  $ & $  5.7 \times 10^{-1}  $ & $  2.3 \times 10^{-1}  $ & $  6.2 \times 10^{-1}  $\\
45--50 & $  8.5 \times 10^{0}  $ & $  6.7 \times 10^{-2}  $ & $  3.7 \times 10^{-1}  $ & $  1.3 \times 10^{-1}  $ & $  4.0 \times 10^{-1}  $\\
50--55 & $  6.3 \times 10^{0}  $ & $  5.5 \times 10^{-2}  $ & $  2.4 \times 10^{-1}  $ & $  1.1 \times 10^{-1}  $ & $  2.7 \times 10^{-1}  $\\
55--60 & $  5.3 \times 10^{0}  $ & $  5.0 \times 10^{-2}  $ & $  2.0 \times 10^{-1}  $ & $  9.8 \times 10^{-2}  $ & $  2.3 \times 10^{-1}  $\\
60--64 & $  4.9 \times 10^{0}  $ & $  5.6 \times 10^{-2}  $ & $  1.7 \times 10^{-1}  $ & $  9.7 \times 10^{-2}  $ & $  2.1 \times 10^{-1}  $\\
64--68 & $  4.9 \times 10^{0}  $ & $  5.8 \times 10^{-2}  $ & $  1.6 \times 10^{-1}  $ & $  1.0 \times 10^{-1}  $ & $  2.0 \times 10^{-1}  $\\
68--72 & $  5.4 \times 10^{0}  $ & $  5.9 \times 10^{-2}  $ & $  1.8 \times 10^{-1}  $ & $  1.1 \times 10^{-1}  $ & $  2.2 \times 10^{-1}  $\\
72--76 & $  6.5 \times 10^{0}  $ & $  6.8 \times 10^{-2}  $ & $  2.0 \times 10^{-1}  $ & $  1.4 \times 10^{-1}  $ & $  2.5 \times 10^{-1}  $\\
76--81 & $  9.7 \times 10^{0}  $ & $  7.0 \times 10^{-2}  $ & $  3.0 \times 10^{-1}  $ & $  2.0 \times 10^{-1}  $ & $  3.7 \times 10^{-1}  $\\
81--86 & $  2.1 \times 10^{1}  $ & $  1.1 \times 10^{-1}  $ & $  6.5 \times 10^{-1}  $ & $  4.2 \times 10^{-1}  $ & $  7.9 \times 10^{-1}  $\\
86--91 & $  1.5 \times 10^{2}  $ & $  2.7 \times 10^{-1}  $ & $  5.0 \times 10^{0}  $ & $  2.9 \times 10^{0}  $ & $  5.8 \times 10^{0}  $\\
91--96 & $  1.6 \times 10^{2}  $ & $  2.7 \times 10^{-1}  $ & $  5.9 \times 10^{0}  $ & $  3.1 \times 10^{0}  $ & $  6.6 \times 10^{0}  $\\
96--101 & $  1.4 \times 10^{1}  $ & $  8.5 \times 10^{-2}  $ & $  5.3 \times 10^{-1}  $ & $  2.4 \times 10^{-1}  $ & $  5.9 \times 10^{-1}  $\\
101--106 & $  4.9 \times 10^{0}  $ & $  4.8 \times 10^{-2}  $ & $  2.0 \times 10^{-1}  $ & $  7.6 \times 10^{-2}  $ & $  2.2 \times 10^{-1}  $\\
106--110 & $  2.5 \times 10^{0}  $ & $  4.3 \times 10^{-2}  $ & $  1.1 \times 10^{-1}  $ & $  3.6 \times 10^{-2}  $ & $  1.2 \times 10^{-1}  $\\
110--115 & $  1.7 \times 10^{0}  $ & $  2.8 \times 10^{-2}  $ & $  7.1 \times 10^{-2}  $ & $  2.2 \times 10^{-2}  $ & $  8.0 \times 10^{-2}  $\\
115--120 & $  1.1 \times 10^{0}  $ & $  2.3 \times 10^{-2}  $ & $  4.7 \times 10^{-2}  $ & $  1.3 \times 10^{-2}  $ & $  5.4 \times 10^{-2}  $\\
120--126 & $  7.6 \times 10^{-1}  $ & $  1.8 \times 10^{-2}  $ & $  3.5 \times 10^{-2}  $ & $  8.6 \times 10^{-3}  $ & $  4.1 \times 10^{-2}  $\\
126--133 & $  5.2 \times 10^{-1}  $ & $  1.3 \times 10^{-2}  $ & $  2.6 \times 10^{-2}  $ & $  5.5 \times 10^{-3}  $ & $  2.9 \times 10^{-2}  $\\
133--141 & $  3.7 \times 10^{-1}  $ & $  1.0 \times 10^{-2}  $ & $  1.9 \times 10^{-2}  $ & $  3.8 \times 10^{-3}  $ & $  2.2 \times 10^{-2}  $\\
141--150 & $  2.7 \times 10^{-1}  $ & $  8.0 \times 10^{-3}  $ & $  1.6 \times 10^{-2}  $ & $  2.7 \times 10^{-3}  $ & $  1.8 \times 10^{-2}  $\\
150--160 & $  1.9 \times 10^{-1}  $ & $  6.1 \times 10^{-3}  $ & $  1.1 \times 10^{-2}  $ & $  1.8 \times 10^{-3}  $ & $  1.3 \times 10^{-2}  $\\
160--171 & $  1.4 \times 10^{-1}  $ & $  5.3 \times 10^{-3}  $ & $  9.1 \times 10^{-3}  $ & $  1.6 \times 10^{-3}  $ & $  1.1 \times 10^{-2}  $\\
171--185 & $  9.1 \times 10^{-2}  $ & $  3.6 \times 10^{-3}  $ & $  6.4 \times 10^{-3}  $ & $  8.0 \times 10^{-4}  $ & $  7.3 \times 10^{-3}  $\\
185--200 & $  6.3 \times 10^{-2}  $ & $  2.9 \times 10^{-3}  $ & $  4.7 \times 10^{-3}  $ & $  6.5 \times 10^{-4}  $ & $  5.6 \times 10^{-3}  $\\
200--220 & $  4.4 \times 10^{-2}  $ & $  1.9 \times 10^{-3}  $ & $  3.6 \times 10^{-3}  $ & $  5.1 \times 10^{-4}  $ & $  4.1 \times 10^{-3}  $\\
220--243 & $  3.0 \times 10^{-2}  $ & $  1.5 \times 10^{-3}  $ & $  2.7 \times 10^{-3}  $ & $  2.7 \times 10^{-4}  $ & $  3.1 \times 10^{-3}  $\\
243--273 & $  1.7 \times 10^{-2}  $ & $  9.8 \times 10^{-4}  $ & $  1.7 \times 10^{-3}  $ & $  1.3 \times 10^{-4}  $ & $  2.0 \times 10^{-3}  $\\
273--320 & $  9.9 \times 10^{-3}  $ & $  5.4 \times 10^{-4}  $ & $  1.1 \times 10^{-3}  $ & $  7.2 \times 10^{-5}  $ & $  1.2 \times 10^{-3}  $\\
320--380 & $  5.3 \times 10^{-3}  $ & $  3.4 \times 10^{-4}  $ & $  6.0 \times 10^{-4}  $ & $  3.4 \times 10^{-5}  $ & $  6.9 \times 10^{-4}  $\\
380--440 & $  1.6 \times 10^{-3}  $ & $  2.3 \times 10^{-4}  $ & $  1.9 \times 10^{-4}  $ & $  1.2 \times 10^{-5}  $ & $  3.0 \times 10^{-4}  $\\
440--510 & $  1.1 \times 10^{-3}  $ & $  1.5 \times 10^{-4}  $ & $  1.3 \times 10^{-4}  $ & $  6.8 \times 10^{-6}  $ & $  2.0 \times 10^{-4}  $\\
510--600 & $  5.7 \times 10^{-4}  $ & $  8.9 \times 10^{-5}  $ & $  7.1 \times 10^{-5}  $ & $  2.6 \times 10^{-6}  $ & $  1.1 \times 10^{-4}  $\\
600--700 & $  3.2 \times 10^{-4}  $ & $  6.2 \times 10^{-5}  $ & $  4.0 \times 10^{-5}  $ & $  9.6 \times 10^{-7}  $ & $  7.4 \times 10^{-5}  $\\
700--830 & $  8.3 \times 10^{-5}  $ & $  3.0 \times 10^{-5}  $ & $  1.8 \times 10^{-5}  $ & $  3.8 \times 10^{-7}  $ & $  3.5 \times 10^{-5}  $\\
830--1000 & $  5.5 \times 10^{-5}  $ & $  1.8 \times 10^{-5}  $ & $  1.1 \times 10^{-5}  $ & $  4.1 \times 10^{-7}  $ & $  2.1 \times 10^{-5}  $\\
1000--1500 & $  1.1 \times 10^{-5}  $ & $  4.1 \times 10^{-6}  $ & $  2.3 \times 10^{-6}  $ & $  1.3 \times 10^{-7}  $ & $  4.7 \times 10^{-6}  $\\
1500--3000 & $  2.4 \times 10^{-7}  $ & $  2.4 \times 10^{-7}  $ & $  3.8 \times 10^{-7}  $ & $  4.2 \times 10^{-9}  $ & $  4.5 \times 10^{-7}  $\\
\hline
\end{tabular}
\end{table}

\begin{table} [htpb]
\centering
\topcaption{
  Summary of the measured values of $\rd\sigma / \rd{m}$ (pb/\GeVns{}) in the dielectron channel with the statistical
($\delta_{\text{stat}}$), experimental ($\delta_{\text{exp}}$)  and theoretical ($\delta_{\text{theo}}$) uncertainties, respectively.
Here, $\delta_{\text{tot}}$ is the quadratic sum of the three components.
}
\label{tab:result-electrons}
\begin{tabular}{ccccccc}
\hline
  $m(\GeVns{})$ & $\frac{\rd\sigma}{\rd{m}}$ (pb/\GeVns{}) & $\delta_{\text{stat}}$ & $\delta_{\text{exp}}$ & $\delta_{\text{theo}}$ & $\delta_{\text{tot}}$ \\
\hline
15--20 & $  2.2 \times 10^{2}  $ & $  5.4 \times 10^{0}  $ & $  1.2 \times 10^{1}  $ & $  1.2 \times 10^{1}  $ & $  1.8 \times 10^{1}  $\\
20--25 & $  1.0 \times 10^{2}  $ & $  3.0 \times 10^{0}  $ & $  5.4 \times 10^{0}  $ & $  4.8 \times 10^{0}  $ & $  7.9 \times 10^{0}  $\\
25--30 & $  5.1 \times 10^{1}  $ & $  1.8 \times 10^{0}  $ & $  2.8 \times 10^{0}  $ & $  4.9 \times 10^{0}  $ & $  5.9 \times 10^{0}  $\\
30--35 & $  2.8 \times 10^{1}  $ & $  1.1 \times 10^{0}  $ & $  1.6 \times 10^{0}  $ & $  4.3 \times 10^{0}  $ & $  4.7 \times 10^{0}  $\\
35--40 & $  1.9 \times 10^{1}  $ & $  7.3 \times 10^{-1}  $ & $  1.3 \times 10^{0}  $ & $  3.0 \times 10^{0}  $ & $  3.3 \times 10^{0}  $\\
40--45 & $  1.1 \times 10^{1}  $ & $  4.6 \times 10^{-1}  $ & $  8.5 \times 10^{-1}  $ & $  1.6 \times 10^{0}  $ & $  1.8 \times 10^{0}  $\\
45--50 & $  8.2 \times 10^{0}  $ & $  3.2 \times 10^{-1}  $ & $  6.3 \times 10^{-1}  $ & $  9.5 \times 10^{-1}  $ & $  1.2 \times 10^{0}  $\\
50--55 & $  5.7 \times 10^{0}  $ & $  2.6 \times 10^{-1}  $ & $  4.2 \times 10^{-1}  $ & $  5.3 \times 10^{-1}  $ & $  7.3 \times 10^{-1}  $\\
55--60 & $  5.7 \times 10^{0}  $ & $  2.1 \times 10^{-1}  $ & $  3.9 \times 10^{-1}  $ & $  4.3 \times 10^{-1}  $ & $  6.1 \times 10^{-1}  $\\
60--64 & $  4.3 \times 10^{0}  $ & $  2.6 \times 10^{-1}  $ & $  2.7 \times 10^{-1}  $ & $  2.8 \times 10^{-1}  $ & $  4.7 \times 10^{-1}  $\\
64--68 & $  4.8 \times 10^{0}  $ & $  2.4 \times 10^{-1}  $ & $  2.7 \times 10^{-1}  $ & $  2.7 \times 10^{-1}  $ & $  4.5 \times 10^{-1}  $\\
68--72 & $  5.5 \times 10^{0}  $ & $  2.7 \times 10^{-1}  $ & $  2.8 \times 10^{-1}  $ & $  2.6 \times 10^{-1}  $ & $  4.7 \times 10^{-1}  $\\
72--76 & $  6.8 \times 10^{0}  $ & $  2.8 \times 10^{-1}  $ & $  3.0 \times 10^{-1}  $ & $  2.7 \times 10^{-1}  $ & $  4.9 \times 10^{-1}  $\\
76--81 & $  1.0 \times 10^{1}  $ & $  2.9 \times 10^{-1}  $ & $  3.8 \times 10^{-1}  $ & $  3.4 \times 10^{-1}  $ & $  5.9 \times 10^{-1}  $\\
81--86 & $  2.4 \times 10^{1}  $ & $  3.8 \times 10^{-1}  $ & $  8.2 \times 10^{-1}  $ & $  6.8 \times 10^{-1}  $ & $  1.1 \times 10^{0}  $\\
86--91 & $  1.5 \times 10^{2}  $ & $  6.0 \times 10^{-1}  $ & $  4.8 \times 10^{0}  $ & $  3.4 \times 10^{0}  $ & $  5.9 \times 10^{0}  $\\
91--96 & $  1.5 \times 10^{2}  $ & $  5.4 \times 10^{-1}  $ & $  5.2 \times 10^{0}  $ & $  3.0 \times 10^{0}  $ & $  6.1 \times 10^{0}  $\\
96--101 & $  1.3 \times 10^{1}  $ & $  1.8 \times 10^{-1}  $ & $  4.7 \times 10^{-1}  $ & $  2.1 \times 10^{-1}  $ & $  5.5 \times 10^{-1}  $\\
101--106 & $  4.9 \times 10^{0}  $ & $  1.0 \times 10^{-1}  $ & $  1.9 \times 10^{-1}  $ & $  6.7 \times 10^{-2}  $ & $  2.2 \times 10^{-1}  $\\
106--110 & $  2.6 \times 10^{0}  $ & $  9.5 \times 10^{-2}  $ & $  1.2 \times 10^{-1}  $ & $  3.2 \times 10^{-2}  $ & $  1.5 \times 10^{-1}  $\\
110--115 & $  1.5 \times 10^{0}  $ & $  6.2 \times 10^{-2}  $ & $  7.7 \times 10^{-2}  $ & $  1.7 \times 10^{-2}  $ & $  1.0 \times 10^{-1}  $\\
115--120 & $  1.1 \times 10^{0}  $ & $  4.8 \times 10^{-2}  $ & $  5.6 \times 10^{-2}  $ & $  1.2 \times 10^{-2}  $ & $  7.5 \times 10^{-2}  $\\
120--126 & $  7.9 \times 10^{-1}  $ & $  3.5 \times 10^{-2}  $ & $  4.2 \times 10^{-2}  $ & $  8.8 \times 10^{-3}  $ & $  5.6 \times 10^{-2}  $\\
126--133 & $  5.7 \times 10^{-1}  $ & $  2.5 \times 10^{-2}  $ & $  3.7 \times 10^{-2}  $ & $  6.6 \times 10^{-3}  $ & $  4.5 \times 10^{-2}  $\\
133--141 & $  3.3 \times 10^{-1}  $ & $  1.9 \times 10^{-2}  $ & $  2.7 \times 10^{-2}  $ & $  3.9 \times 10^{-3}  $ & $  3.3 \times 10^{-2}  $\\
141--150 & $  3.1 \times 10^{-1}  $ & $  1.6 \times 10^{-2}  $ & $  2.7 \times 10^{-2}  $ & $  3.6 \times 10^{-3}  $ & $  3.1 \times 10^{-2}  $\\
150--160 & $  1.7 \times 10^{-1}  $ & $  1.1 \times 10^{-2}  $ & $  1.5 \times 10^{-2}  $ & $  1.9 \times 10^{-3}  $ & $  1.9 \times 10^{-2}  $\\
160--171 & $  1.4 \times 10^{-1}  $ & $  8.9 \times 10^{-3}  $ & $  1.3 \times 10^{-2}  $ & $  1.4 \times 10^{-3}  $ & $  1.6 \times 10^{-2}  $\\
171--185 & $  1.0 \times 10^{-1}  $ & $  6.5 \times 10^{-3}  $ & $  9.8 \times 10^{-3}  $ & $  9.0 \times 10^{-4}  $ & $  1.2 \times 10^{-2}  $\\
185--200 & $  5.4 \times 10^{-2}  $ & $  4.4 \times 10^{-3}  $ & $  5.4 \times 10^{-3}  $ & $  4.6 \times 10^{-4}  $ & $  6.9 \times 10^{-3}  $\\
200--220 & $  4.3 \times 10^{-2}  $ & $  3.2 \times 10^{-3}  $ & $  4.3 \times 10^{-3}  $ & $  3.8 \times 10^{-4}  $ & $  5.4 \times 10^{-3}  $\\
220--243 & $  3.0 \times 10^{-2}  $ & $  2.4 \times 10^{-3}  $ & $  3.2 \times 10^{-3}  $ & $  2.2 \times 10^{-4}  $ & $  4.0 \times 10^{-3}  $\\
243--273 & $  1.5 \times 10^{-2}  $ & $  1.5 \times 10^{-3}  $ & $  1.6 \times 10^{-3}  $ & $  1.0 \times 10^{-4}  $ & $  2.2 \times 10^{-3}  $\\
273--320 & $  9.3 \times 10^{-3}  $ & $  7.9 \times 10^{-4}  $ & $  9.9 \times 10^{-4}  $ & $  6.0 \times 10^{-5}  $ & $  1.3 \times 10^{-3}  $\\
320--380 & $  4.9 \times 10^{-3}  $ & $  5.0 \times 10^{-4}  $ & $  5.2 \times 10^{-4}  $ & $  2.8 \times 10^{-5}  $ & $  7.2 \times 10^{-4}  $\\
380--440 & $  2.8 \times 10^{-3}  $ & $  3.6 \times 10^{-4}  $ & $  3.2 \times 10^{-4}  $ & $  1.3 \times 10^{-5}  $ & $  4.8 \times 10^{-4}  $\\
440--510 & $  5.0 \times 10^{-4}  $ & $  1.9 \times 10^{-4}  $ & $  6.4 \times 10^{-5}  $ & $  2.0 \times 10^{-6}  $ & $  2.0 \times 10^{-4}  $\\
510--600 & $  5.9 \times 10^{-4}  $ & $  1.2 \times 10^{-4}  $ & $  8.3 \times 10^{-5}  $ & $  2.2 \times 10^{-6}  $ & $  1.4 \times 10^{-4}  $\\
600--700 & $  3.2 \times 10^{-4}  $ & $  8.2 \times 10^{-5}  $ & $  5.6 \times 10^{-5}  $ & $  1.0 \times 10^{-6}  $ & $  1.0 \times 10^{-4}  $\\
700--830 & $  5.9 \times 10^{-5}  $ & $  3.5 \times 10^{-5}  $ & $  1.4 \times 10^{-5}  $ & $  1.7 \times 10^{-7}  $ & $  3.8 \times 10^{-5}  $\\
830--1000 & $  9.6 \times 10^{-6}  $ & $  5.6 \times 10^{-6}  $ & $  4.3 \times 10^{-6}  $ & $  2.5 \times 10^{-8}  $ & $  7.1 \times 10^{-6}  $\\
1000--1500 & $  9.1 \times 10^{-6}  $ & $  5.0 \times 10^{-6}  $ & $  7.5 \times 10^{-6}  $ & $  3.3 \times 10^{-8}  $ & $  9.0 \times 10^{-6}  $\\
1500--3000 & $  4.3 \times 10^{-7}  $ & $  4.3 \times 10^{-7}  $ & $  4.4 \times 10^{-7}  $ & $  2.8 \times 10^{-9}  $ & $  6.2 \times 10^{-7}  $\\
\hline
\end{tabular}
\end{table}

\begin{figure}[htpb]
{\centering
\includegraphics[width=0.75\textwidth]{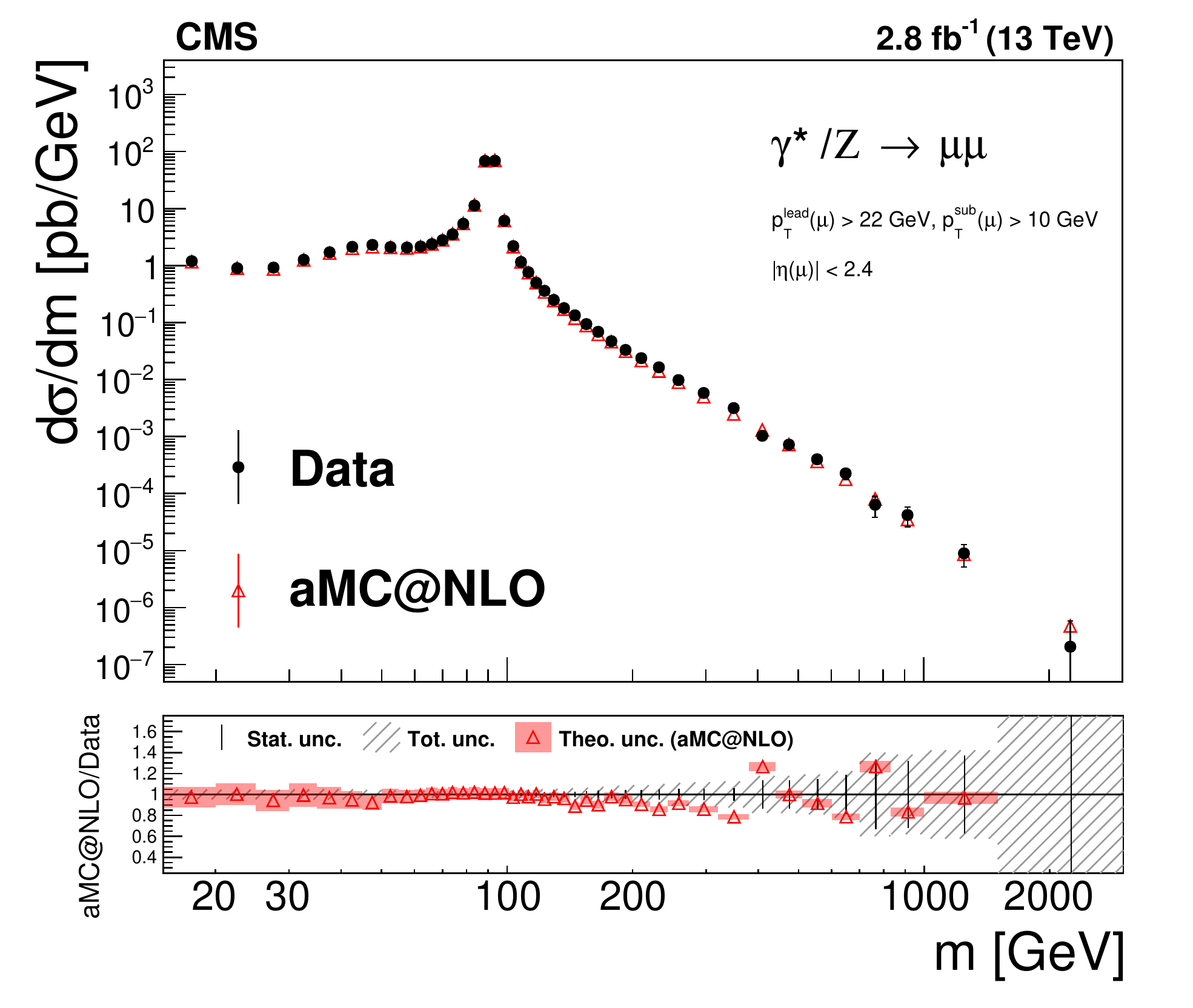}
\includegraphics[width=0.75\textwidth]{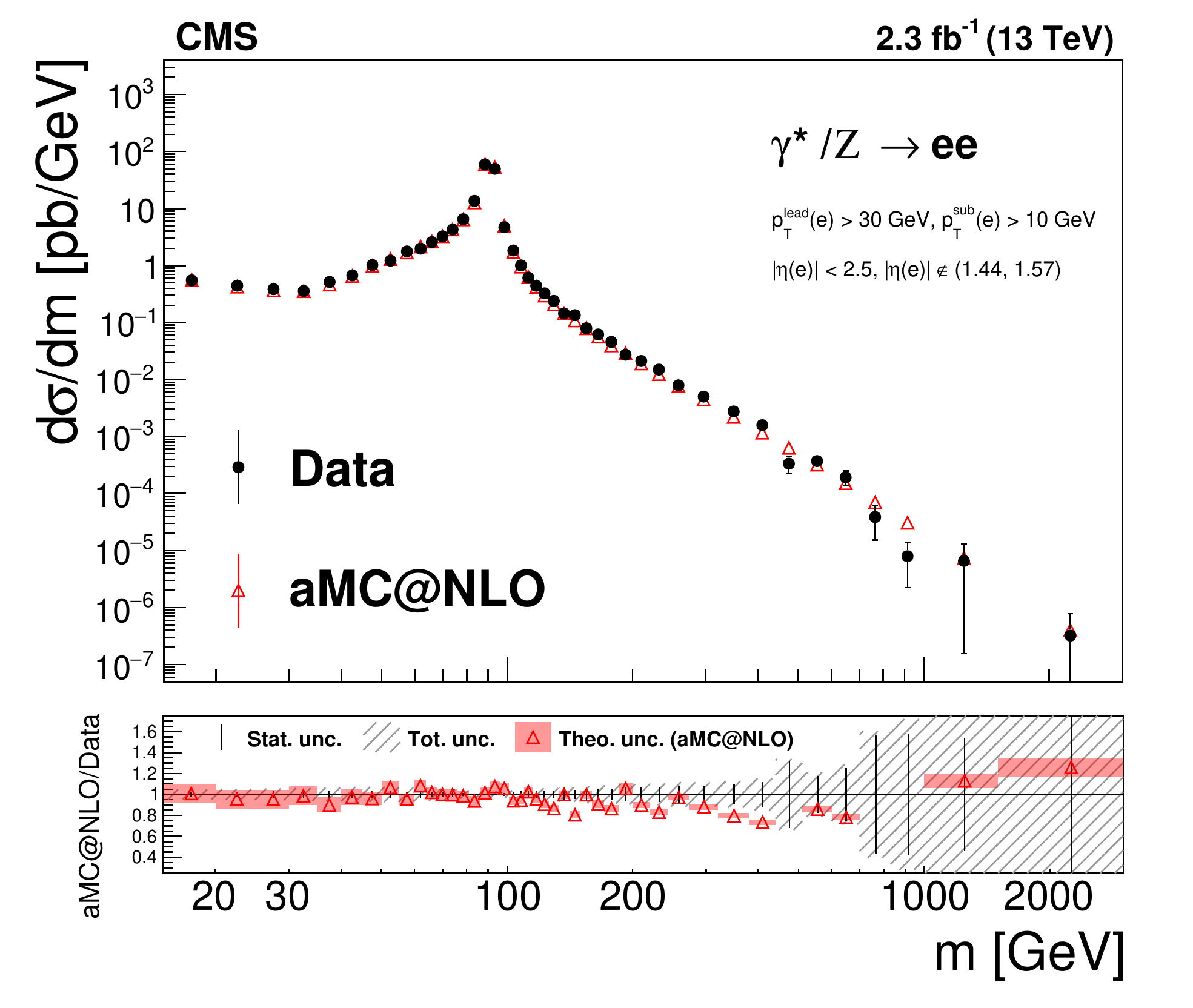}
\caption{
  \label{fig:dsdM_mcnlo_fiducial}
  Comparison between the measured fiducial cross section (with no FSR correction applied)
  and the NLO theoretical prediction of \MGvATNLO with NNPDF 3.0 in the dimuon (upper) and dielectron (lower) channels.
  In the bottom panels, the red band represents the theoretical uncertainty and
  the hatched band represents the total uncertainty, which is the combination of the statistical,
  systematic, and integrated luminosity components.
}
}
\end{figure}

In addition to the fully corrected total cross section, the fiducial cross section is also measured within the detector acceptance and without the FSR correction.
Because of the fact that the acceptance correction, as shown in Fig.~\ref{fig:1Dacceff},
is very large below the \cPZ~boson peak, the shape of the fiducial distribution is
different in this mass region.
Figure~\ref{fig:dsdM_mcnlo_fiducial} shows the results in the dimuon and dielectron channels compared to the prediction from \MGvATNLO. Tables~\ref{tab:result-muons_FpoF} and~\ref{tab:result-electrons_FpoF} present the summary of the fiducial cross section measurements (with no FSR correction applied) for the dimuon and dielectron channels, respectively.
The results are in good agreement, within the uncertainties, with the theoretical prediction.

\begin{table} [htpb]
\centering
\topcaption{Summary of the measured values of fiducial $\rd\sigma / \rd{m}$ (pb/\GeVns{}) (with no FSR correction applied) in the dimuon channel with the statistical
($\delta_{\text{stat}}$) and experimental ($\delta_{\text{exp}}$) uncertainties shown separately.
Here, $\delta_{\text{tot}}$ is the quadratic sum of the two components.
}
\label{tab:result-muons_FpoF}
\begin{tabular}{cccccc}
\hline
  $m(\GeVns{})$ & $\frac{\rd\sigma}{\rd{m}}$ (pb/\GeVns{}) & $\delta_{\text{stat}}$ & $\delta_{\text{exp}}$ & $\delta_{\text{tot}}$ \\
\hline
15--20 & $  1.2 \times 10^{0}  $ & $  1.1 \times 10^{-2}  $ & $  5.5 \times 10^{-2}  $ & $  5.6 \times 10^{-2}  $\\
20--25 & $  9.0 \times 10^{-1}  $ & $  9.8 \times 10^{-3}  $ & $  4.0 \times 10^{-2}  $ & $  4.2 \times 10^{-2}  $\\
25--30 & $  9.2 \times 10^{-1}  $ & $  1.1 \times 10^{-2}  $ & $  4.2 \times 10^{-2}  $ & $  4.3 \times 10^{-2}  $\\
30--35 & $  1.3 \times 10^{0}  $ & $  1.3 \times 10^{-2}  $ & $  6.1 \times 10^{-2}  $ & $  6.3 \times 10^{-2}  $\\
35--40 & $  1.7 \times 10^{0}  $ & $  1.5 \times 10^{-2}  $ & $  8.6 \times 10^{-2}  $ & $  8.7 \times 10^{-2}  $\\
40--45 & $  2.1 \times 10^{0}  $ & $  1.7 \times 10^{-2}  $ & $  1.0 \times 10^{-1}  $ & $  1.0 \times 10^{-1}  $\\
45--50 & $  2.3 \times 10^{0}  $ & $  1.7 \times 10^{-2}  $ & $  1.0 \times 10^{-1}  $ & $  1.0 \times 10^{-1}  $\\
50--55 & $  2.1 \times 10^{0}  $ & $  1.7 \times 10^{-2}  $ & $  8.1 \times 10^{-2}  $ & $  8.2 \times 10^{-2}  $\\
55--60 & $  2.1 \times 10^{0}  $ & $  1.7 \times 10^{-2}  $ & $  7.5 \times 10^{-2}  $ & $  7.7 \times 10^{-2}  $\\
60--64 & $  2.2 \times 10^{0}  $ & $  2.0 \times 10^{-2}  $ & $  7.2 \times 10^{-2}  $ & $  7.5 \times 10^{-2}  $\\
64--68 & $  2.4 \times 10^{0}  $ & $  2.2 \times 10^{-2}  $ & $  7.2 \times 10^{-2}  $ & $  7.6 \times 10^{-2}  $\\
68--72 & $  2.8 \times 10^{0}  $ & $  2.3 \times 10^{-2}  $ & $  8.1 \times 10^{-2}  $ & $  8.4 \times 10^{-2}  $\\
72--76 & $  3.5 \times 10^{0}  $ & $  2.7 \times 10^{-2}  $ & $  9.6 \times 10^{-2}  $ & $  9.9 \times 10^{-2}  $\\
76--81 & $  5.4 \times 10^{0}  $ & $  2.8 \times 10^{-2}  $ & $  1.4 \times 10^{-1}  $ & $  1.5 \times 10^{-1}  $\\
81--86 & $  1.1 \times 10^{1}  $ & $  4.5 \times 10^{-2}  $ & $  3.1 \times 10^{-1}  $ & $  3.1 \times 10^{-1}  $\\
86--91 & $  6.8 \times 10^{1}  $ & $  1.1 \times 10^{-1}  $ & $  2.0 \times 10^{0}  $ & $  2.0 \times 10^{0}  $\\
91--96 & $  6.9 \times 10^{1}  $ & $  1.1 \times 10^{-1}  $ & $  2.2 \times 10^{0}  $ & $  2.2 \times 10^{0}  $\\
96--101 & $  6.1 \times 10^{0}  $ & $  3.6 \times 10^{-2}  $ & $  2.1 \times 10^{-1}  $ & $  2.1 \times 10^{-1}  $\\
101--106 & $  2.2 \times 10^{0}  $ & $  2.0 \times 10^{-2}  $ & $  7.8 \times 10^{-2}  $ & $  8.1 \times 10^{-2}  $\\
106--110 & $  1.2 \times 10^{0}  $ & $  1.9 \times 10^{-2}  $ & $  4.3 \times 10^{-2}  $ & $  4.7 \times 10^{-2}  $\\
110--115 & $  7.6 \times 10^{-1}  $ & $  1.2 \times 10^{-2}  $ & $  3.0 \times 10^{-2}  $ & $  3.2 \times 10^{-2}  $\\
115--120 & $  5.0 \times 10^{-1}  $ & $  1.0 \times 10^{-2}  $ & $  2.1 \times 10^{-2}  $ & $  2.3 \times 10^{-2}  $\\
120--126 & $  3.6 \times 10^{-1}  $ & $  7.9 \times 10^{-3}  $ & $  1.6 \times 10^{-2}  $ & $  1.8 \times 10^{-2}  $\\
126--133 & $  2.5 \times 10^{-1}  $ & $  5.8 \times 10^{-3}  $ & $  1.2 \times 10^{-2}  $ & $  1.3 \times 10^{-2}  $\\
133--141 & $  1.8 \times 10^{-1}  $ & $  4.5 \times 10^{-3}  $ & $  8.9 \times 10^{-3}  $ & $  1.0 \times 10^{-2}  $\\
141--150 & $  1.3 \times 10^{-1}  $ & $  3.7 \times 10^{-3}  $ & $  7.2 \times 10^{-3}  $ & $  8.1 \times 10^{-3}  $\\
150--160 & $  9.4 \times 10^{-2}  $ & $  2.9 \times 10^{-3}  $ & $  5.3 \times 10^{-3}  $ & $  6.1 \times 10^{-3}  $\\
160--171 & $  6.9 \times 10^{-2}  $ & $  2.5 \times 10^{-3}  $ & $  4.2 \times 10^{-3}  $ & $  4.9 \times 10^{-3}  $\\
171--185 & $  4.7 \times 10^{-2}  $ & $  1.7 \times 10^{-3}  $ & $  3.1 \times 10^{-3}  $ & $  3.6 \times 10^{-3}  $\\
185--200 & $  3.3 \times 10^{-2}  $ & $  1.4 \times 10^{-3}  $ & $  2.4 \times 10^{-3}  $ & $  2.8 \times 10^{-3}  $\\
200--220 & $  2.4 \times 10^{-2}  $ & $  9.7 \times 10^{-4}  $ & $  1.9 \times 10^{-3}  $ & $  2.1 \times 10^{-3}  $\\
220--243 & $  1.6 \times 10^{-2}  $ & $  7.7 \times 10^{-4}  $ & $  1.4 \times 10^{-3}  $ & $  1.6 \times 10^{-3}  $\\
243--273 & $  9.8 \times 10^{-3}  $ & $  5.3 \times 10^{-4}  $ & $  9.4 \times 10^{-4}  $ & $  1.1 \times 10^{-3}  $\\
273--320 & $  5.8 \times 10^{-3}  $ & $  3.1 \times 10^{-4}  $ & $  6.0 \times 10^{-4}  $ & $  6.7 \times 10^{-4}  $\\
320--380 & $  3.2 \times 10^{-3}  $ & $  2.0 \times 10^{-4}  $ & $  3.4 \times 10^{-4}  $ & $  3.9 \times 10^{-4}  $\\
380--440 & $  1.0 \times 10^{-3}  $ & $  1.4 \times 10^{-4}  $ & $  1.2 \times 10^{-4}  $ & $  1.8 \times 10^{-4}  $\\
440--510 & $  7.2 \times 10^{-4}  $ & $  9.6 \times 10^{-5}  $ & $  8.5 \times 10^{-5}  $ & $  1.3 \times 10^{-4}  $\\
510--600 & $  4.0 \times 10^{-4}  $ & $  6.0 \times 10^{-5}  $ & $  4.8 \times 10^{-5}  $ & $  7.6 \times 10^{-5}  $\\
600--700 & $  2.3 \times 10^{-4}  $ & $  4.3 \times 10^{-5}  $ & $  2.7 \times 10^{-5}  $ & $  5.0 \times 10^{-5}  $\\
700--830 & $  6.4 \times 10^{-5}  $ & $  2.1 \times 10^{-5}  $ & $  1.4 \times 10^{-5}  $ & $  2.5 \times 10^{-5}  $\\
830--1000 & $  4.2 \times 10^{-5}  $ & $  1.4 \times 10^{-5}  $ & $  8.4 \times 10^{-6}  $ & $  1.6 \times 10^{-5}  $\\
1000--1500 & $  8.9 \times 10^{-6}  $ & $  3.3 \times 10^{-6}  $ & $  1.8 \times 10^{-6}  $ & $  3.8 \times 10^{-6}  $\\
1500--3000 & $  2.1 \times 10^{-7}  $ & $  2.1 \times 10^{-7}  $ & $  3.2 \times 10^{-7}  $ & $  3.8 \times 10^{-7}  $\\
\hline
\end{tabular}
\end{table}

\begin{table} [htpb]
\centering
\topcaption{Summary of the measured values of fiducial $\rd\sigma / \rd{m}$ (pb/\GeVns{}) (with no FSR correction applied) in the dielectron channel with the statistical
($\delta_{\text{stat}}$) and experimental ($\delta_{\text{exp}}$) uncertainties shown separately.
Here, $\delta_{\text{tot}}$ is the quadratic sum of the two components.
}
\label{tab:result-electrons_FpoF}
\begin{tabular}{cccccc}
\hline
  $m(\GeVns{})$ & $\frac{\rd\sigma}{\rd{m}}$ (pb/\GeVns{}) & $\delta_{\text{stat}}$ & $\delta_{\text{exp}}$ & $\delta_{\text{tot}}$ \\
\hline
15--20 & $  5.5 \times 10^{-1}  $ & $  1.3 \times 10^{-2}  $ & $  2.3 \times 10^{-2}  $ & $  2.6 \times 10^{-2}  $\\
20--25 & $  4.4 \times 10^{-1}  $ & $  1.2 \times 10^{-2}  $ & $  1.9 \times 10^{-2}  $ & $  2.3 \times 10^{-2}  $\\
25--30 & $  3.9 \times 10^{-1}  $ & $  1.2 \times 10^{-2}  $ & $  1.7 \times 10^{-2}  $ & $  2.1 \times 10^{-2}  $\\
30--35 & $  3.6 \times 10^{-1}  $ & $  1.3 \times 10^{-2}  $ & $  1.7 \times 10^{-2}  $ & $  2.2 \times 10^{-2}  $\\
35--40 & $  5.1 \times 10^{-1}  $ & $  1.8 \times 10^{-2}  $ & $  2.9 \times 10^{-2}  $ & $  3.4 \times 10^{-2}  $\\
40--45 & $  6.7 \times 10^{-1}  $ & $  2.3 \times 10^{-2}  $ & $  4.1 \times 10^{-2}  $ & $  4.7 \times 10^{-2}  $\\
45--50 & $  1.0 \times 10^{0}  $ & $  3.1 \times 10^{-2}  $ & $  6.7 \times 10^{-2}  $ & $  7.4 \times 10^{-2}  $\\
50--55 & $  1.2 \times 10^{0}  $ & $  3.8 \times 10^{-2}  $ & $  8.0 \times 10^{-2}  $ & $  8.9 \times 10^{-2}  $\\
55--60 & $  1.8 \times 10^{0}  $ & $  4.3 \times 10^{-2}  $ & $  1.1 \times 10^{-1}  $ & $  1.2 \times 10^{-1}  $\\
60--64 & $  2.0 \times 10^{0}  $ & $  6.0 \times 10^{-2}  $ & $  1.2 \times 10^{-1}  $ & $  1.4 \times 10^{-1}  $\\
64--68 & $  2.6 \times 10^{0}  $ & $  6.7 \times 10^{-2}  $ & $  1.4 \times 10^{-1}  $ & $  1.6 \times 10^{-1}  $\\
68--72 & $  3.2 \times 10^{0}  $ & $  7.6 \times 10^{-2}  $ & $  1.6 \times 10^{-1}  $ & $  1.8 \times 10^{-1}  $\\
72--76 & $  4.3 \times 10^{0}  $ & $  8.3 \times 10^{-2}  $ & $  1.9 \times 10^{-1}  $ & $  2.0 \times 10^{-1}  $\\
76--81 & $  6.5 \times 10^{0}  $ & $  8.7 \times 10^{-2}  $ & $  2.4 \times 10^{-1}  $ & $  2.5 \times 10^{-1}  $\\
81--86 & $  1.4 \times 10^{1}  $ & $  1.2 \times 10^{-1}  $ & $  4.3 \times 10^{-1}  $ & $  4.5 \times 10^{-1}  $\\
86--91 & $  5.9 \times 10^{1}  $ & $  1.9 \times 10^{-1}  $ & $  1.7 \times 10^{0}  $ & $  1.7 \times 10^{0}  $\\
91--96 & $  5.0 \times 10^{1}  $ & $  1.7 \times 10^{-1}  $ & $  1.4 \times 10^{0}  $ & $  1.4 \times 10^{0}  $\\
96--101 & $  4.7 \times 10^{0}  $ & $  5.9 \times 10^{-2}  $ & $  1.4 \times 10^{-1}  $ & $  1.5 \times 10^{-1}  $\\
101--106 & $  1.8 \times 10^{0}  $ & $  3.4 \times 10^{-2}  $ & $  5.8 \times 10^{-2}  $ & $  6.7 \times 10^{-2}  $\\
106--110 & $  1.0 \times 10^{0}  $ & $  3.1 \times 10^{-2}  $ & $  4.0 \times 10^{-2}  $ & $  5.1 \times 10^{-2}  $\\
110--115 & $  6.1 \times 10^{-1}  $ & $  2.1 \times 10^{-2}  $ & $  2.8 \times 10^{-2}  $ & $  3.5 \times 10^{-2}  $\\
115--120 & $  4.4 \times 10^{-1}  $ & $  1.7 \times 10^{-2}  $ & $  2.1 \times 10^{-2}  $ & $  2.7 \times 10^{-2}  $\\
120--126 & $  3.3 \times 10^{-1}  $ & $  1.3 \times 10^{-2}  $ & $  1.5 \times 10^{-2}  $ & $  1.9 \times 10^{-2}  $\\
126--133 & $  2.4 \times 10^{-1}  $ & $  9.3 \times 10^{-3}  $ & $  1.1 \times 10^{-2}  $ & $  1.4 \times 10^{-2}  $\\
133--141 & $  1.4 \times 10^{-1}  $ & $  7.0 \times 10^{-3}  $ & $  7.0 \times 10^{-3}  $ & $  9.9 \times 10^{-3}  $\\
141--150 & $  1.3 \times 10^{-1}  $ & $  5.9 \times 10^{-3}  $ & $  7.0 \times 10^{-3}  $ & $  9.2 \times 10^{-3}  $\\
150--160 & $  7.9 \times 10^{-2}  $ & $  4.4 \times 10^{-3}  $ & $  4.4 \times 10^{-3}  $ & $  6.3 \times 10^{-3}  $\\
160--171 & $  6.2 \times 10^{-2}  $ & $  3.5 \times 10^{-3}  $ & $  3.7 \times 10^{-3}  $ & $  5.1 \times 10^{-3}  $\\
171--185 & $  4.6 \times 10^{-2}  $ & $  2.6 \times 10^{-3}  $ & $  3.0 \times 10^{-3}  $ & $  3.9 \times 10^{-3}  $\\
185--200 & $  2.7 \times 10^{-2}  $ & $  1.9 \times 10^{-3}  $ & $  1.9 \times 10^{-3}  $ & $  2.6 \times 10^{-3}  $\\
200--220 & $  2.1 \times 10^{-2}  $ & $  1.4 \times 10^{-3}  $ & $  1.5 \times 10^{-3}  $ & $  2.1 \times 10^{-3}  $\\
220--243 & $  1.5 \times 10^{-2}  $ & $  1.1 \times 10^{-3}  $ & $  1.1 \times 10^{-3}  $ & $  1.6 \times 10^{-3}  $\\
243--273 & $  7.9 \times 10^{-3}  $ & $  6.7 \times 10^{-4}  $ & $  6.4 \times 10^{-4}  $ & $  9.3 \times 10^{-4}  $\\
273--320 & $  5.0 \times 10^{-3}  $ & $  4.0 \times 10^{-4}  $ & $  4.2 \times 10^{-4}  $ & $  5.8 \times 10^{-4}  $\\
320--380 & $  2.8 \times 10^{-3}  $ & $  2.6 \times 10^{-4}  $ & $  2.4 \times 10^{-4}  $ & $  3.5 \times 10^{-4}  $\\
380--440 & $  1.6 \times 10^{-3}  $ & $  1.9 \times 10^{-4}  $ & $  1.5 \times 10^{-4}  $ & $  2.4 \times 10^{-4}  $\\
440--510 & $  3.4 \times 10^{-4}  $ & $  1.1 \times 10^{-4}  $ & $  3.6 \times 10^{-5}  $ & $  1.1 \times 10^{-4}  $\\
510--600 & $  3.7 \times 10^{-4}  $ & $  6.6 \times 10^{-5}  $ & $  4.3 \times 10^{-5}  $ & $  7.9 \times 10^{-5}  $\\
600--700 & $  1.9 \times 10^{-4}  $ & $  4.9 \times 10^{-5}  $ & $  2.7 \times 10^{-5}  $ & $  5.5 \times 10^{-5}  $\\
700--830 & $  3.9 \times 10^{-5}  $ & $  2.2 \times 10^{-5}  $ & $  7.6 \times 10^{-6}  $ & $  2.3 \times 10^{-5}  $\\
830--1000 & $  8.0 \times 10^{-6}  $ & $  4.6 \times 10^{-6}  $ & $  3.4 \times 10^{-6}  $ & $  5.7 \times 10^{-6}  $\\
1000--1500 & $  6.6 \times 10^{-6}  $ & $  3.5 \times 10^{-6}  $ & $  5.4 \times 10^{-6}  $ & $  6.4 \times 10^{-6}  $\\
1500--3000 & $  3.2 \times 10^{-7}  $ & $  3.2 \times 10^{-7}  $ & $  3.2 \times 10^{-7}  $ & $  4.6 \times 10^{-7}  $\\

\hline
\end{tabular}
\end{table}

The differential cross section measurements obtained in the dimuon and dielectron channels in the full phase space are
in good agreement with each other and therefore the results in the two channels are combined using the method described in Ref.~\cite{bib:ZJets8TeV}.
The combined differential cross section in each bin is determined by the average of the two measurements weighted by the inverse of the squared total uncertainty.
The uncertainty of the combined result is extracted from the diagonal elements of the covariance matrix of the combination,
which is constructed using the covariance matrices of the measurements in the dimuon and dielectron channels.
The uncertainties are considered to be uncorrelated between the two channels, with
the exception of the integrated luminosity and the acceptance uncertainties.
Systematic uncertainties are treated as additional uncorrelated uncertainties in the individual measurements, with the exception of the uncertainty from the efficiency scale factor.
Statistical uncertainties are propagated to the covariance matrices of the measurements in the dimuon and
dielectron channels before the combination.

\begin{figure}[htpb]
{\centering
\includegraphics[width=0.75\textwidth]{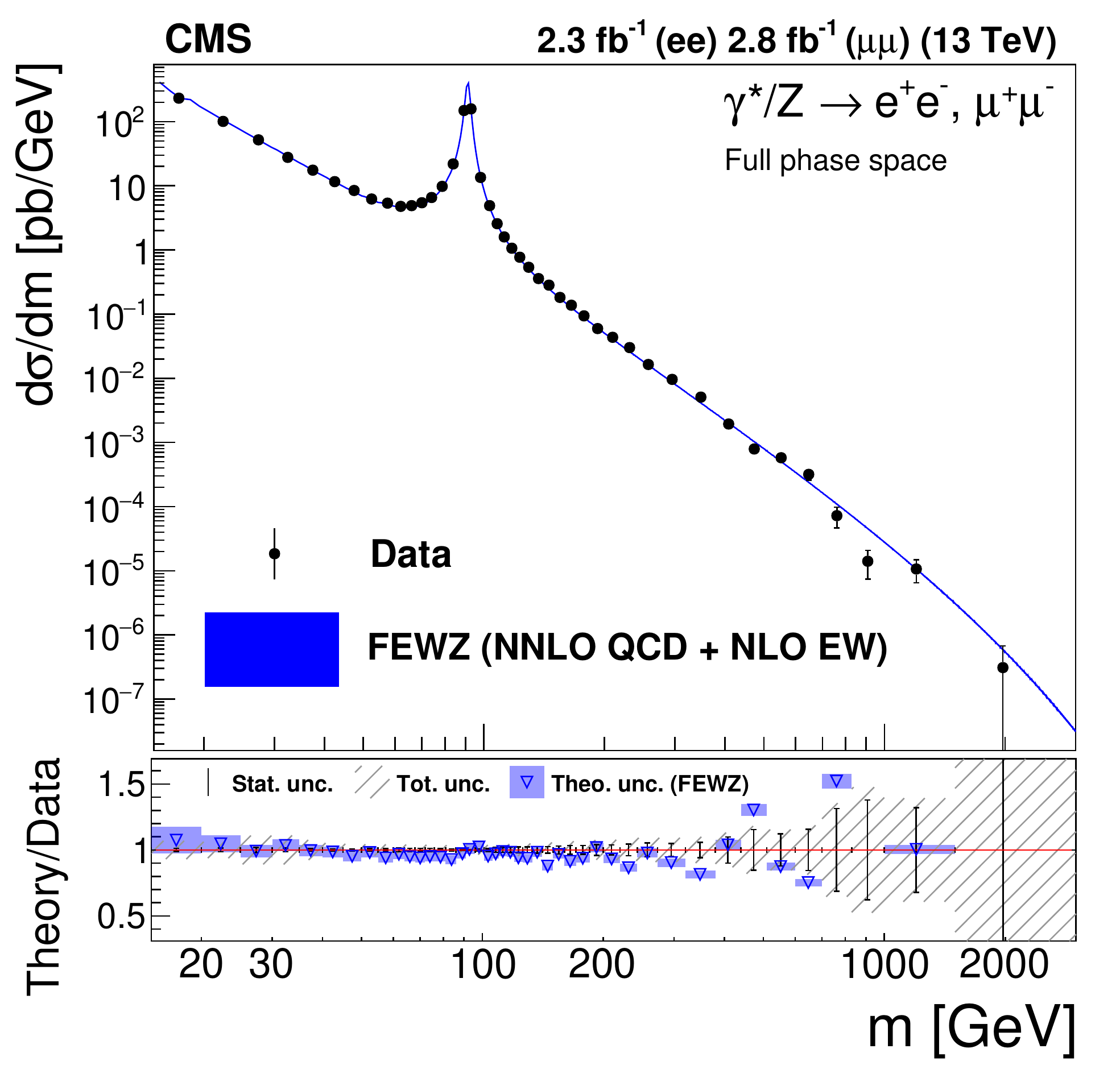}
\caption{
  \label{fig:dsdM_combined}
The differential DY cross section measured for the combination of the two channels and as predicted
by the NNLO theoretical calculation of \FEWZ with NNPDF 3.0 in the full phase space.
The ratio between the data and the theoretical prediction is presented in the bottom panel.
The coloured boxes represent the theoretical uncertainties.
}
}
\end{figure}

\begin{figure}[htpb]
{\centering
\includegraphics[width=0.67\textwidth]{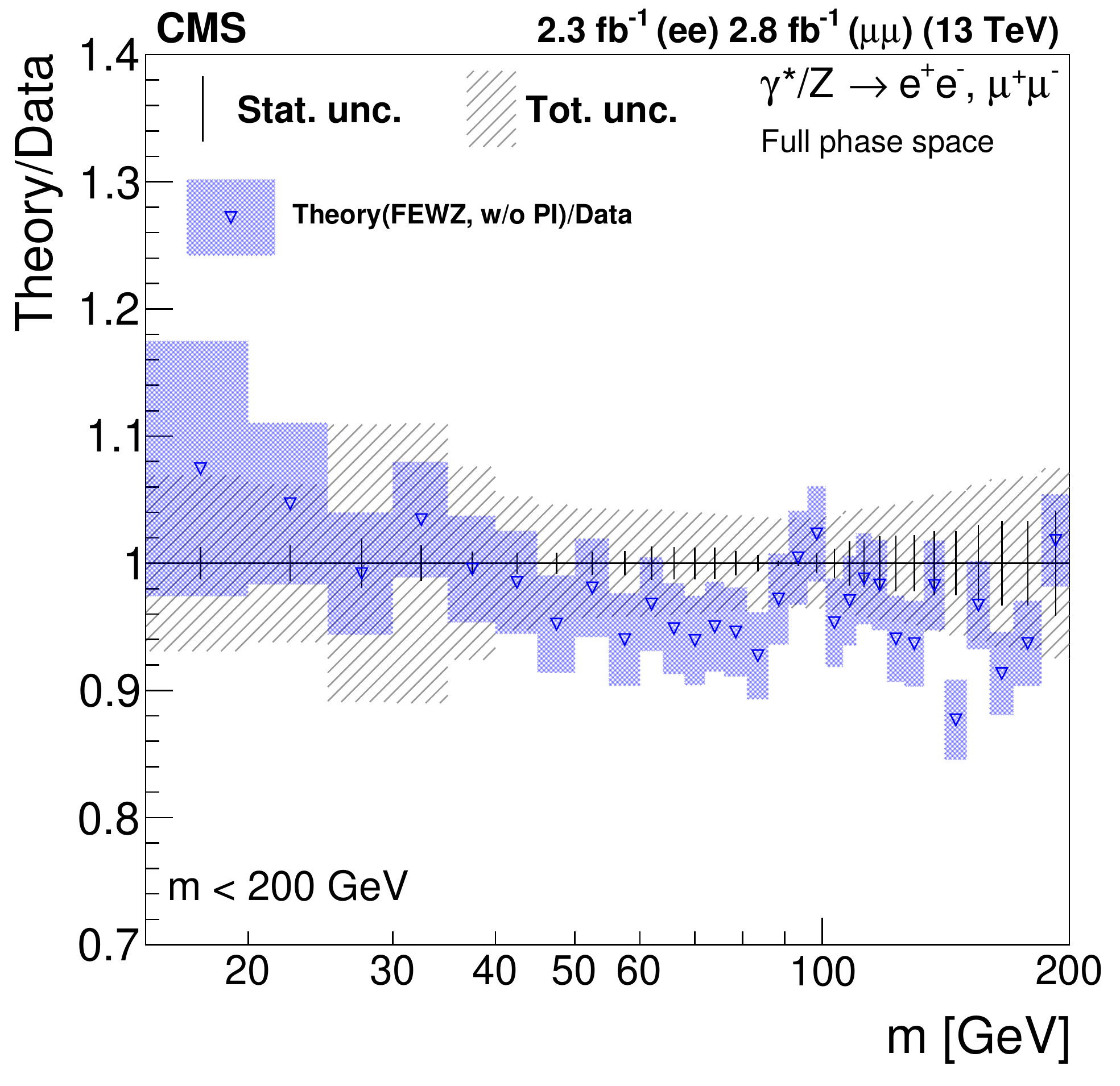}
\includegraphics[width=0.67\textwidth]{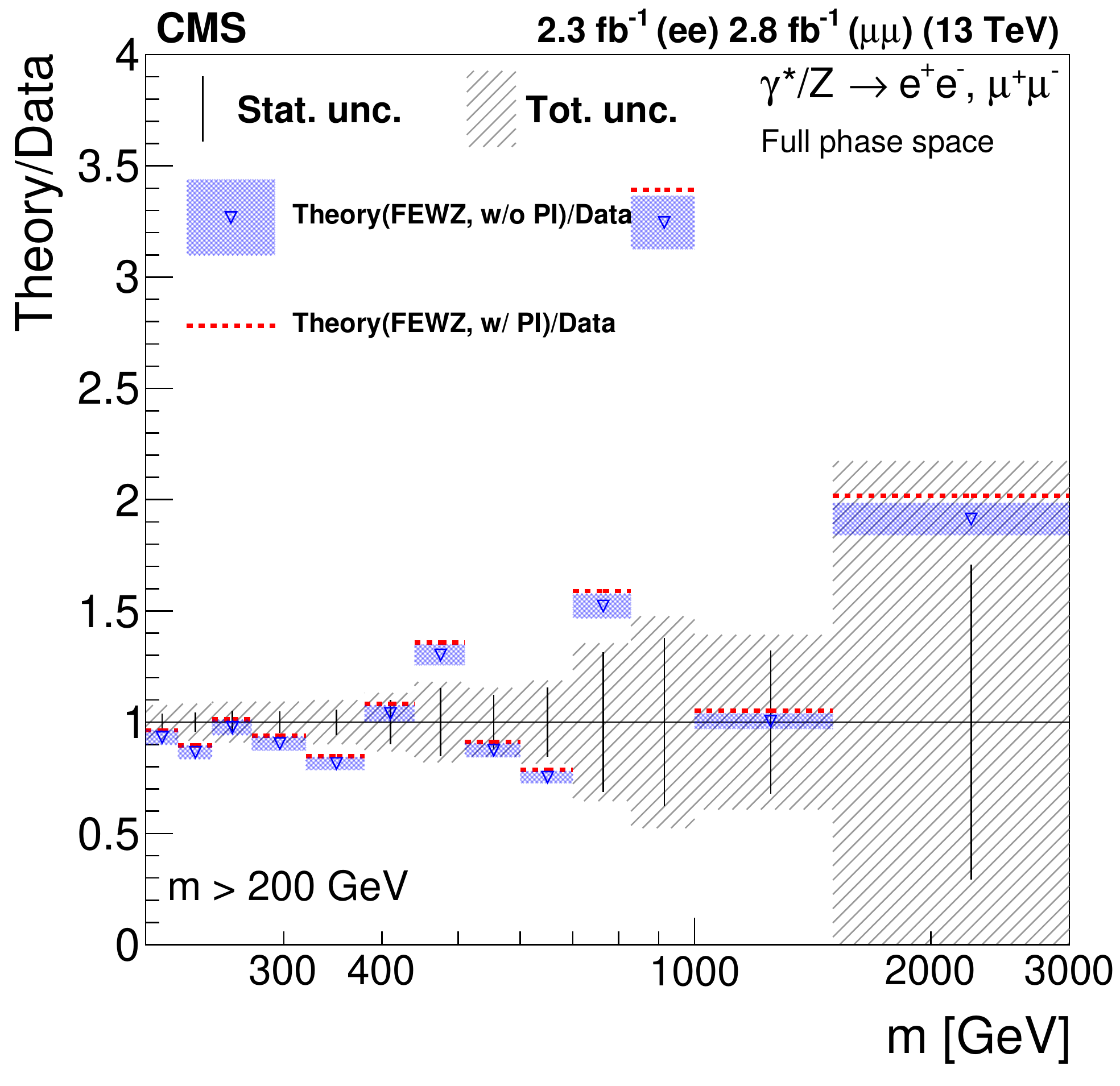}
\caption{
  \label{fig:comb_ratio}
Magnified view of the ratio of the NNLO theoretical prediction of \FEWZ with NNPDF 3.0 to data for the
combined differential cross sections in two different mass ranges:
$m < 200\GeV$ (top) and $m > 200\GeV$ (bottom).
The blue bands represent the theoretical uncertainty on the ratio.
The bottom plot also shows the ratio with the photon-induced contribution (red dashed lines), which has a sizable effect in the high-mass region.
}
}
\end{figure}

The differential DY cross section for the full phase space, after the combination of the dimuon and dielectron channels, is presented in Fig.~\ref{fig:dsdM_combined}.
The data point abscissas are computed according to Eq. (6) in Ref.~\cite{bib:pts_plac}.
Table~\ref{tab:result-comb} shows the summary of the combined results.
Figure~\ref{fig:comb_ratio} shows a magnified ratio plot for the comparison of the theoretical prediction with experimental result in two mass ranges, $15<m<200\GeV$ and $200<m<3000\GeV$.
In the bottom plot of the figure, an additional theoretical prediction,
containing the photon-induced contribution calculated using \FEWZ with
the LUXqed PDF set, is included. There is a sizable effect from this
contribution in the high-mass region.
The level of agreement between data and theory is very good as indicated by the $p$-value of 0.49 except for $830<m<1000\GeV$, determined from the $\chi^2$ comparison between the combined result and NNLO prediction.

\begin{table} [htpb]
\centering
\topcaption{Summary of the combined values of $\rd\sigma / \rd{m}$ (pb/\GeVns{}) using the results from
both the dimuon and dielectron channels.
Here, $\delta_{\text{tot}}$ is the quadratic sum of the statistical, experimental and theoretical uncertainties.
}
\label{tab:result-comb}
\begin{tabular}{cccc}
\hline
  $m(\GeVns{})$ & $\frac{\rd\sigma}{\rd{m}}$ (pb/\GeVns{}) & $\delta_{\text{tot}}$ \\
\hline
15--20 & $  2.3 \times 10^{2}  $ & $  1.6 \times 10^{1}  $\\
20--25 & $  1.0 \times 10^{2}  $ & $  6.3 \times 10^{0}  $\\
25--30 & $  5.2 \times 10^{1}  $ & $  5.7 \times 10^{0}  $\\
30--35 & $  2.8 \times 10^{1}  $ & $  3.0 \times 10^{0}  $\\
35--40 & $  1.8 \times 10^{1}  $ & $  1.3 \times 10^{0}  $\\
40--45 & $  1.2 \times 10^{1}  $ & $  6.1 \times 10^{-1}  $\\
45--50 & $  8.5 \times 10^{0}  $ & $  3.9 \times 10^{-1}  $\\
50--55 & $  6.2 \times 10^{0}  $ & $  2.7 \times 10^{-1}  $\\
55--60 & $  5.4 \times 10^{0}  $ & $  2.3 \times 10^{-1}  $\\
60--64 & $  4.8 \times 10^{0}  $ & $  2.0 \times 10^{-1}  $\\
64--68 & $  4.9 \times 10^{0}  $ & $  2.0 \times 10^{-1}  $\\
68--72 & $  5.4 \times 10^{0}  $ & $  2.2 \times 10^{-1}  $\\
72--76 & $  6.6 \times 10^{0}  $ & $  2.6 \times 10^{-1}  $\\
76--81 & $  9.8 \times 10^{0}  $ & $  3.7 \times 10^{-1}  $\\
81--86 & $  2.2 \times 10^{1}  $ & $  7.9 \times 10^{-1}  $\\
86--91 & $  1.5 \times 10^{2}  $ & $  5.2 \times 10^{0}  $\\
91--96 & $  1.6 \times 10^{2}  $ & $  5.6 \times 10^{0}  $\\
96--101 & $  1.3 \times 10^{1}  $ & $  4.8 \times 10^{-1}  $\\
101--106 & $  4.9 \times 10^{0}  $ & $  1.8 \times 10^{-1}  $\\
106--110 & $  2.6 \times 10^{0}  $ & $  1.1 \times 10^{-1}  $\\
110--115 & $  1.6 \times 10^{0}  $ & $  6.8 \times 10^{-2}  $\\
115--120 & $  1.1 \times 10^{0}  $ & $  4.7 \times 10^{-2}  $\\
120--126 & $  7.7 \times 10^{-1}  $ & $  3.5 \times 10^{-2}  $\\
126--133 & $  5.4 \times 10^{-1}  $ & $  2.6 \times 10^{-2}  $\\
133--141 & $  3.6 \times 10^{-1}  $ & $  1.9 \times 10^{-2}  $\\
141--150 & $  2.8 \times 10^{-1}  $ & $  1.6 \times 10^{-2}  $\\
150--160 & $  1.8 \times 10^{-1}  $ & $  1.1 \times 10^{-2}  $\\
160--171 & $  1.4 \times 10^{-1}  $ & $  9.1 \times 10^{-3}  $\\
171--185 & $  9.4 \times 10^{-2}  $ & $  6.4 \times 10^{-3}  $\\
185--200 & $  6.0 \times 10^{-2}  $ & $  4.5 \times 10^{-3}  $\\
200--220 & $  4.4 \times 10^{-2}  $ & $  3.3 \times 10^{-3}  $\\
220--243 & $  3.0 \times 10^{-2}  $ & $  2.5 \times 10^{-3}  $\\
243--273 & $  1.6 \times 10^{-2}  $ & $  1.5 \times 10^{-3}  $\\
273--320 & $  9.6 \times 10^{-3}  $ & $  8.9 \times 10^{-4}  $\\
320--380 & $  5.1 \times 10^{-3}  $ & $  5.0 \times 10^{-4}  $\\
380--440 & $  1.9 \times 10^{-3}  $ & $  2.5 \times 10^{-4}  $\\
440--510 & $  7.9 \times 10^{-4}  $ & $  1.4 \times 10^{-4}  $\\
510--600 & $  5.8 \times 10^{-4}  $ & $  8.9 \times 10^{-5}  $\\
600--700 & $  3.2 \times 10^{-4}  $ & $  6.0 \times 10^{-5}  $\\
700--830 & $  7.2 \times 10^{-5}  $ & $  2.6 \times 10^{-5}  $\\
830--1000 & $  1.4 \times 10^{-5}  $ & $  6.7 \times 10^{-6}  $\\
1000--1500 & $  1.1 \times 10^{-5}  $ & $  4.2 \times 10^{-6}  $\\
1500--3000 & $  3.1 \times 10^{-7}  $ & $  3.6 \times 10^{-7}  $\\
\hline
\end{tabular}
\end{table}

\section{Summary}
\label{sec:summary}
This paper presents measurements of the total and fiducial
Drell--Yan differential cross sections $\rd\sigma / \rd{m}$ in the dimuon and the dielectron channels
as well as their combination, in the dilepton invariant mass range $15 < m < 3000\GeV$, using data collected by the CMS
experiment, in proton-proton collisions at a centre-of-mass energy of 13\TeV, corresponding to an
integrated luminosity of up to 2.8\fbinv. The data are corrected for detector
resolution effects, the differences in the efficiency between data and Monte Carlo simulation, and the acceptance.
Additionally the final-state photon radiation effects, which are most pronounced below the \cPZ~boson peak,
are included.
The results are in good agreement with the theoretical predictions of the standard model.
\begin{acknowledgments}
We congratulate our colleagues in the CERN accelerator departments for the excellent performance of the LHC and thank the technical and administrative staffs at CERN and at other CMS institutes for their contributions to the success of the CMS effort. In addition, we gratefully acknowledge the computing centres and personnel of the Worldwide LHC Computing Grid for delivering so effectively the computing infrastructure essential to our analyses. Finally, we acknowledge the enduring support for the construction and operation of the LHC and the CMS detector provided by the following funding agencies: BMBWF and FWF (Austria); FNRS and FWO (Belgium); CNPq, CAPES, FAPERJ, FAPERGS, and FAPESP (Brazil); MES (Bulgaria); CERN; CAS, MoST, and NSFC (China); COLCIENCIAS (Colombia); MSES and CSF (Croatia); RPF (Cyprus); SENESCYT (Ecuador); MoER, ERC IUT, and ERDF (Estonia); Academy of Finland, MEC, and HIP (Finland); CEA and CNRS/IN2P3 (France); BMBF, DFG, and HGF (Germany); GSRT (Greece); NKFIA (Hungary); DAE and DST (India); IPM (Iran); SFI (Ireland); INFN (Italy); MSIP and NRF (Republic of Korea); MES (Latvia); LAS (Lithuania); MOE and UM (Malaysia); BUAP, CINVESTAV, CONACYT, LNS, SEP, and UASLP-FAI (Mexico); MOS (Montenegro); MBIE (New Zealand); PAEC (Pakistan); MSHE and NSC (Poland); FCT (Portugal); JINR (Dubna); MON, RosAtom, RAS, RFBR, and NRC KI (Russia); MESTD (Serbia); SEIDI, CPAN, PCTI, and FEDER (Spain); MOSTR (Sri Lanka); Swiss Funding Agencies (Switzerland); MST (Taipei); ThEPCenter, IPST, STAR, and NSTDA (Thailand); TUBITAK and TAEK (Turkey); NASU and SFFR (Ukraine); STFC (United Kingdom); DOE and NSF (USA).

  \hyphenation{Rachada-pisek} Individuals have received support from the Marie-Curie programme and the European Research Council and Horizon 2020 Grant, contract No. 675440 (European Union); the Leventis Foundation; the A. P. Sloan Foundation; the Alexander von Humboldt Foundation; the Belgian Federal Science Policy Office; the Fonds pour la Formation \`a la Recherche dans l'Industrie et dans l'Agriculture (FRIA-Belgium); the Agentschap voor Innovatie door Wetenschap en Technologie (IWT-Belgium); the F.R.S.-FNRS and FWO (Belgium) under the ``Excellence of Science - EOS" - be.h project n. 30820817; the Ministry of Education, Youth and Sports (MEYS) of the Czech Republic; the Lend\"ulet (``Momentum") Programme and the J\'anos Bolyai Research Scholarship of the Hungarian Academy of Sciences, the New National Excellence Program \'UNKP, the NKFIA research grants 123842, 123959, 124845, 124850 and 125105 (Hungary); the Council of Science and Industrial Research, India; the HOMING PLUS programme of the Foundation for Polish Science, cofinanced from European Union, Regional Development Fund, the Mobility Plus programme of the Ministry of Science and Higher Education, the National Science Center (Poland), contracts Harmonia 2014/14/M/ST2/00428, Opus 2014/13/B/ST2/02543, 2014/15/B/ST2/03998, and 2015/19/B/ST2/02861, Sonata-bis 2012/07/E/ST2/01406; the National Priorities Research Program by Qatar National Research Fund; the Programa Estatal de Fomento de la Investigaci{\'o}n Cient{\'i}fica y T{\'e}cnica de Excelencia Mar\'{\i}a de Maeztu, grant MDM-2015-0509 and the Programa Severo Ochoa del Principado de Asturias; the Thalis and Aristeia programmes cofinanced by EU-ESF and the Greek NSRF; the Rachadapisek Sompot Fund for Postdoctoral Fellowship, Chulalongkorn University and the Chulalongkorn Academic into Its 2nd Century Project Advancement Project (Thailand); the Welch Foundation, contract C-1845; and the Weston Havens Foundation (USA).
\end{acknowledgments}
\bibliography{auto_generated}
\cleardoublepage \appendix\section{The CMS Collaboration \label{app:collab}}\begin{sloppypar}\hyphenpenalty=5000\widowpenalty=500\clubpenalty=5000\vskip\cmsinstskip
\textbf{Yerevan Physics Institute, Yerevan, Armenia}\\*[0pt]
A.M.~Sirunyan, A.~Tumasyan
\vskip\cmsinstskip
\textbf{Institut f\"{u}r Hochenergiephysik, Wien, Austria}\\*[0pt]
W.~Adam, F.~Ambrogi, E.~Asilar, T.~Bergauer, J.~Brandstetter, E.~Brondolin, M.~Dragicevic, J.~Er\"{o}, A.~Escalante~Del~Valle, M.~Flechl, R.~Fr\"{u}hwirth\cmsAuthorMark{1}, V.M.~Ghete, J.~Hrubec, M.~Jeitler\cmsAuthorMark{1}, N.~Krammer, I.~Kr\"{a}tschmer, D.~Liko, T.~Madlener, I.~Mikulec, N.~Rad, H.~Rohringer, J.~Schieck\cmsAuthorMark{1}, R.~Sch\"{o}fbeck, M.~Spanring, D.~Spitzbart, A.~Taurok, W.~Waltenberger, J.~Wittmann, C.-E.~Wulz\cmsAuthorMark{1}, M.~Zarucki
% ...
\vskip\cmsinstskip
% ...
1:  Also at Vienna University of Technology, Vienna, Austria\\
% ...
\end{sloppypar}
\end{document}